\documentstyle{JFM}
\newcommand{\be}{\begin{equation}}
\newcommand{\ee}{\end{equation}}
\newcommand{\bea}{\begin{eqnarray}}
\newcommand{\eea}{\end{eqnarray}}

\newcommand{\vx}{{\bf x}}
\newcommand{\vX}{{\bf X}}
\newcommand{\vv}{{\bf v}}

\newcommand{\hkj}{{\hat{k}_j}}
\newcommand{\hkl}{{\hat{k}_l}}

\newcommand{\vk}{{\bf k}}

\newcommand{\nablatwo}{{\nabla_{\bot}}}

\newcommand{\te}{{\stackrel{\leftrightarrow}{\bf e}}}
\newcommand{\ttau}{{\stackrel{\leftrightarrow}{\mbox{\boldmath $\tau$}}}}

\newcommand{\sF}{{\cal F}}

\newcommand{\sL}{{\cal L}}

\newcommand{\sD}{{\hat{{\cal D}}}}
\newcommand{\sG}{{\hat{{\cal G}}}}

\input{psfig}

\title[Pattern formation in weakly damped Faraday waves]
{Pattern formation in weakly damped parametric surface waves}
\author[W. Zhang and J. Vi\~nals]{Wenbin Zhang$^1$ and Jorge Vi\~nals$^2$}
\affiliation{$^1$Department of Chemical Engineering, 
             Massachusetts Institute of Technology, Cambridge, 
             Massachusetts 02139, USA\\
             $^2$Supercomputer Computations Research Institute,
             Florida State University, Tallahassee, Florida 32306-4052, USA, 
             and Department of Chemical Engineering, 
             FAMU-FSU College of Engineering, Tallahassee, Florida 31310-6046, 
USA}
\pubyear{1996}
\volume{000}
\pagerange{000--000}
\date{\today}
\begin{document}
\maketitle

\begin{abstract}

We present a theoretical study of nonlinear pattern formation in parametric 
surface waves for fluids of low viscosity, and in the limit of large aspect 
ratio.  The analysis is based on a quasi-potential approximation 
to the equations governing fluid motion, followed by a multiscale asymptotic 
expansion in the distance away from threshold.  Close to onset, the 
asymptotic expansion yields an amplitude equation which is of gradient form,
and allows the explicit calculation of the functional form of the cubic 
nonlinearities.
In particular,  we find that three-wave resonant interactions contribute
significantly to the nonlinear terms, and therefore are important 
for pattern selection. Minimization of the associated Lyapunov functional 
predicts 
a primary bifurcation to a standing wave pattern of square symmetry for 
capillary-dominated surface waves, in agreement with experiments.  In 
addition, we find that patterns of hexagonal and quasi-crystalline symmetry 
can be stabilized in certain mixed capillary-gravity waves, even in this
case of single frequency forcing. Quasi-crystalline patterns are predicted
in a region of parameters readily accessible experimentally.

\end{abstract}

\section{Introduction}
\label{sec:introduction}
 
The generation of standing waves on the free surface of a fluid
layer that is oscillated vertically is known since the work of 
\cite{re:faraday31}.  Recently, there has been renewed experimental interest 
in Faraday waves as an example of a pattern-forming system. Reasons
include the ease of experimentation due to short characteristic time scales 
(of the order of $10^{-2}$ seconds),  and the ability to reach very large 
aspect ratios (the ratio of lateral size of the system to the 
characteristic wavelength of the pattern)
of the order of $10^2$.  By varying the form of the driving force 
and by using fluids of different viscosities, a number of interesting 
phenomena have been observed including the emergence of standing wave 
patterns of 
different symmetries near onset (Christiansen, Alstr{\o~}m \& Levinsen 1992;
Fauve {\it et al.} 1992; Edwards \& Fauve 1993, 1994; M\"uller 1993),
secondary instabilities of these patterns when the amplitude of the 
periodic driving force is increased (Ezerskii, Korotin \& Rabinovich 1985;
Daudet {\it et al.} 1995), and spatiotemporal chaotic states at even larger 
amplitudes of the driving force (Tufillaro, Ramshankar \& Gollub 1989;
Gollub \& Ramshankar 1991; Bosch \& van de Water 1993; 
Bridger {\it et al.} 1993; Kudrolli \& Gollub 1996).  

A numerical linear stability analysis of the Faraday wave problem has been 
carried out by \cite{re:kumar94} for a laterally infinite fluid layer of 
arbitrary viscosity.  The predicted values of the acceleration
threshold and the wavelength at onset are in good agreement with 
experiments in large aspect ratio systems
(see also Bechhoefer {\it et al.} 1995).  In the simpler case of
an ideal fluid, a classical linear stability analysis leads 
to the Mathieu equation for the interface displacement (Benjamin \& Ursell 
1954). 
On the other hand, the nonlinear evolution above onset, including 
pattern selection, secondary instabilities and the transition to 
spatiotemporal chaos are not very well understood theoretically.
In this paper, we present a weakly 
nonlinear analysis of Faraday waves driven by a sinusoidal force in the 
limit of weak dissipation and for a laterally infinite system of
infinite depth (or unbounded free surface waves).  Our main focus is on 
pattern selection  near onset,  and our results are compared to
experiments involving fluids of low viscosity, in containers
of large depth and aspect ratio. 

Many studies of free surface waves in incompressible fluids have 
focused on the inviscid limit (see, for example, Yuen \& Lake 1982;
Craik 1985).  Viscous dissipation, however, is essential for Faraday waves 
because it not only sets a threshold value of the driving force, but also
affects nonlinear saturation of the parametric instability.      
For unbounded free surface waves in the limit of weak dissipation, 
the flow remains potential except in a very thin layer at the free surface
(Lamb 1932; Landau \& Lifshitz 1959).  A common procedure
in this limit involves the introduction of the so-called quasi-potential
approximation (QPA) which perturbatively incorporates weak viscous effects by 
introducing modified boundary conditions for the otherwise potential bulk 
flow. By performing a formal expansion in the small thickness of the viscous
boundary layer, Lundgren \& Mansour (1988) derived a set of 
quasi-potential equations (QPE's) that included nonlinear viscous 
contributions 
for the free surface flow of an axially symmetric liquid drop.  
Ruvinsky, Feldstein \& Freidman (1991) later derived a set of QPE's that 
contain 
only linear damping terms for two-dimensional free surface waves.
In~\S\,\ref{sec:equations}, we present QPE's for parametric surface
waves including only linear viscous terms.
These equations are a direct extension of those of Ruvinsky, Feldstein \& 
Freidman for the two-dimensional case.  They can also be derived in 
a formal expansion similar to that of Lundgren \& Mansour (1988) and neglecting
nonlinear viscous terms.  Such formal derivation and the expressions for
the nonlinear viscous terms can be found in Zhang (1994).  

Since we are interested in the surface displacement, but not in the 
flow field in the fluid interior, additional simplification is 
achieved by writing the three-dimensional QPE's in a two-dimensional nonlocal 
form, that involves flow variables at the free surface alone.  
Such simplification is possible because the velocity potential satisfies 
Laplace's equation in the fluid interior, and is determined uniquely
when its values on the boundary (the free surface) are known.  An outline
of the derivation of the {\em weakly nonlinear} two-dimensional form of the 
QPE's is given in~\S\,\ref{sec:2DQPE} by using the so-called 
Dirichlet-Neumann operator (Craig 1989).  The projection of the 
fluid equations onto equations involving the coordinates of the free 
surface follows the works of \cite{re:miles77}, \cite{re:milder77}, 
and \cite{re:craig89} for inviscid gravity waves.
These two-dimensional nonlocal quasi-potential equations are our starting 
point for our weakly nonlinear analysis of parametric surface waves which is
presented in~\S\,\ref{sec:sinusoidal}.  Standing wave amplitude equations are
derived  by using a multiple scale perturbation expansion.   Pattern selection
in Faraday waves near onset is discussed in~\S\,\ref{sec:SWAE-selection}.

Although the relation between our results and earlier theoretical work on
Faraday waves is discussed in more detail below, we note here 
two important studies addressing the weakly nonlinear regime above onset
by Milner (1991), and Miles (1993, 1994).  
They obtained amplitude equations for inviscid flow, and then
introduced weak viscous effects by adding damping terms directly to the
amplitude equations from an energy balance consideration. The resulting
amplitude equations differ from ours in several respects, hence our 
results differ qualitatively from theirs, as discussed 
in~\S\,\ref{sec:sinusoidal}. 

\section{Quasi-potential equations for free surface waves}
\label{sec:equations}

We consider a reference state in which a quiescent and incompressible 
Newtonian liquid of density $\rho$ and kinematic viscosity $\nu$ occupies 
the $z<0$ half space, and a gas of negligible density occupies the $z>0$ 
half space.  The gas phase is also assumed to have a uniform and
constant pressure field $p_0$.  Under these conditions, the velocity
distribution of the gas phase may be allowed to remain unknown and we
can write the governing equations for the liquid phase only. 
Such a liquid-gas interface is usually called the 
{\em free\/} surface of a liquid (Batchelor 1967).
The governing equations for the velocity field $\vv(x,y,z,t)$ 
are for $ -\infty < z < h(x,y,t)$, where $z=h(x,y,t)$ is the instantaneous
location of the free surface,
\bea
\nabla \cdot \vv = 0,   \label{eq:incomp} \\
\partial_{t} \vv + (\vv \cdot \nabla) \vv = - \frac{1}{\rho}\nabla p +
\nu\nabla^{2}\vv + g(t)\hat{z},         \label{eq:ns_orig}
\eea
where $\hat{z}$ is the unit vector in the positive $z$ direction,
and $g(t) = -g_0 - g_z(t)$ with $g_0$ a constant gravitational acceleration,
and $g_z(t)$ the effective acceleration caused by
the vertical oscillation of the fluid in the Faraday experiments.  
The boundary conditions at the free surface are,
\bea
\partial_{t}h/\sqrt{1+(\nabla h)^2} = \vv \cdot \hat{n}, \label{eq:kinbound} \\
\hat{a} \cdot \ttau \cdot \hat{n} \equiv
2\rho\nu\hat{a} \cdot \te \cdot \hat{n} = 0,    \label{eq:tang1bound} \\
\hat{b} \cdot \ttau \cdot \hat{n} \equiv
2\rho\nu\hat{b} \cdot \te \cdot \hat{n} = 0,        \label{eq:tang2bound} \\
p - \hat{n} \cdot \ttau  \cdot\hat{n} \equiv
p - 2\rho\nu\hat{n}\cdot  \te \cdot\hat{n} = p_{0} + \Gamma \kappa,
\label{eq:dynabound}
\eea
with boundary condition,
\be
\mbox{$\vv = 0$, as $z \rightarrow -\infty$}, \label{eq:bottbound}
\ee
where $\hat{a}(x,y,t)$ and $\hat{b}(x,y,t)$ are two tangential unit
vectors on the free surface, and
$\hat{n}(x,y,t)=(-h_{x},-h_{y}, 1)/\sqrt{1+(\nabla h)^2}$
is the unit normal vector of the free surface pointing away from the 
liquid.  $\Gamma$ is the surface tension,
$\kappa$ is the mean curvature of the free surface, which is given by
$\kappa = \nabla \cdot \hat{n}$,
and $\ttau = 2\rho\nu\te$ is the viscous stress tensor, where $\te$ is
the rate of strain tensor with Cartesian components ($i,j=x,y,z$)
\be
(\te)_{ij} = \frac{1}{2}\left(\frac{\partial v_{i}}{\partial x_{j}} +
                          \frac{\partial v_{j}}{\partial x_{i}}\right).
\label{eq:strain_tensor}
\ee

The equations governing fluid motion can be simplified in the limit of 
weak viscous dissipation. In this case,
a thin viscous boundary layer, also known as the vortical layer,
occurs near the free surface as a result of
the nonzero irrotational shear stresses.  This small
irrotational shear stress drags a thin viscous layer of rotational fluid 
along, causing a small modification in the velocity field which is required
in order to satisfy the zero-shear-stress boundary conditions 
(Eqs.~(\ref{eq:tang1bound}) and (\ref{eq:tang2bound})). 
Since the free surface vortical layer is
thin, the basic idea of the quasi-potential approximation is to
consider pure potential flow in
the bulk that satisfies effective boundary conditions on the moving surface
to account for weak viscous effects.

\subsection{Three-dimensional form of the quasi-potential equations}
\label{sec:3DQPE}

Let $\phi(x,y,z,t)$ be the velocity potential for the bulk potential flow.
As a direct extension of the QPE's of Ruvinsky, Feldstein \& Freidman (1991)
for two-dimensional surface waves, the governing equations
for unbounded three-dimensional surface waves read, 
\be
\label{eq:laplace_phi}
\nabla^{2}\phi(x,y,z,t) = 0, \>\>\>\>\>\> \mbox{for} \>\>\>\>\>\> z < h(x,y,t),
\ee
with boundary conditions at the free surface $z=h(x,y,t)$
\bea
\label{eq:kinb_qp}
\partial_{t}h + \nabla\phi \cdot \nabla h = \partial_{z}\phi + W(x,y,t), 
\\
\label{eq:dynb_qp}
\partial_{t}  + \frac{1}{2}(\nabla\phi)^{2}
-g(t)h + 2\nu\frac{\partial^{2}\phi}{\partial z^{2}}
= \Gamma \kappa,
\\
\label{eq:W_qb}
\partial_{t}W(x,y,t) = 2\nu\left(\partial_{xx}+\partial_{yy}\right)
                       \partial_z\phi,
\eea
and
\be
\label{eq:bc_infty}
\partial_{z}\phi \rightarrow 0, \>\>\>\>\>\> \mbox{as} \>\>\>\>\>\> z
\rightarrow -\infty,
\ee
where $W(x,y,t)$ is the $z$-component (or the linearized normal component) 
of the rotational part of the velocity field at the free surface.  The
viscous contribution in Eq.~(\ref{eq:dynb_qp}) is related to
the normal stress of the irrotational velocity component. 
We note that only linear viscous (damping) terms are retained in the above 
set of equations, which we shall refer to as LDQPE's.
Nonlinear viscous contributions can be obtained by using a formal perturbation
expansion, similar to that of Lundgren \& Mansour (1988), of 
Eq.~(\ref{eq:incomp}-\ref{eq:bottbound}) in the small thickness 
of the viscous boundary layer at the free
surface.  Details of the expansion can be found in Zhang (1994).  
These nonlinear viscous terms for three-dimensional waves are, 
however, algebraically too complicated to be included in our analysis. 
We note that neglecting nonlinear viscous contributions in the above equations
is an uncontrolled approximation, motivated by the small viscosity of the 
fluid. Whether the LDQPE's are a good approximation for weakly 
damped parametric surface waves near onset is one of the central issues 
of this paper.  Based on our analytical results of 
the LDQPE's in~\S\,\ref{sec:sinusoidal}, and the comparison of these 
results with experiments, we conclude that the LDQPE's do provide 
a quite good description of weakly damped and weakly nonlinear Faraday waves.
This seems to indicate that the role of nonlinear viscous terms in the QPE's 
is not significant for pattern formation in weakly damped Faraday waves close 
to 
onset.

The $z$-component of the rotational part of the velocity $W(x,y,t)$ can 
be eliminated. From Eqs.~(\ref{eq:kinb_qp}) and (\ref{eq:W_qb}), we have,
\be
\partial_{t}W = 2\nu(\partial_{xx}+\partial_{yy}) \partial_t h
              + 2\nu(\partial_{xx}+\partial_{yy})
                \left(\nabla\phi\cdot\nabla h - W\right).
\ee
We note that $\nu\nabla\phi\cdot\nabla h$ is a nonlinear viscous term,
and thus negligible within the approximation.
Since $W$ is of ${\cal O}(\nu)$, the term $\nu W$ in the above equation
is of ${\cal O}(\nu^{-2})$, and is negligible in the weakly damped limit.  
Therefore we have $\partial_{t}(W-2\nu\nabla^2 h) = 0$, or
\be
W(x,y,t) = 2\nu\nabla^2 h(x,y,t) + W_0 - 2\nu\nabla^2 h_0,
\ee
where $h_0$ and $W_0$ are initial conditions for $h$ and $W$ respectively. 
By considering fluid motion starting from at rest, we can set $W_0 = 0$.
Also since we are interested in nonlinear pattern formation 
from a nearly flat surface, $h_0$ is a small quantity.   Although the 
term $2\nu\nabla^2 h_0$ might influence the linear growth,  it is
certainly negligible for nonlinear, finite amplitude states.  
Thus the boundary conditions at $z =  h(x,y,t)$ (Eqs. 
(\ref{eq:kinb_qp}) and (\ref{eq:dynb_qp})) now read,
\bea
\label{eq:boundary1}
\partial_t h = 2 \nu\nabla^2 h - \nablatwo\phi \cdot \nabla h 
             + \partial_z \phi,
\\
\label{eq:boundary2}
\partial_t \phi = 2 \nu\nablatwo^2 \phi 
                - \frac{1}{2}\left(\nabla\phi\right)^2
                + g(t) h - \Gamma\kappa,
\eea
where $\nablatwo = \partial_{x} \hat{x} + \partial_{y} \hat{y}$.

\subsection{Two-dimensional nonlocal form of the quasi-potential equations}
\label{sec:2DQPE}
For weakly nonlinear surface waves, the three-dimensional quasi-potential
equations can be further simplified by recasting them in a form that
involves only the flow variables on the free surface.
Since the velocity potential $\phi$ satisfies Laplace's equation in the bulk,
it is possible to rewrite the LDQPE's as integro-differential equations 
involving variables at the free surface only.  We then expand the resulting
equations to third order in the wave steepness.
Such a two-dimensional nonlocal formulation has been derived by
\cite{re:miles77}, \cite{re:milder77}, \cite{re:craig89}, and 
\cite{re:craig93} for 
un-forced inviscid gravity waves.  We extend their approach in this section to
parametrically forced, weakly damped capillary-gravity waves.

Let $\vx = (x,y)$, and define the surface velocity potential $\Phi(\vx,t)$ as
$\Phi(\vx,t) = \phi(\vx, h(\vx,t), t)$.
Then Eqs. (\ref{eq:boundary1}) and (\ref{eq:boundary2}) can be rewritten
as,
\bea
\partial_t h = 2 \nu \nabla^2 h + \sqrt{1+(\nabla h)^2} \partial_n\phi,
\label{eq:bcx1}
\\
\partial_t \Phi = 2 \nu \nabla^2 \Phi 
                - \frac{1}{2}\left(\nabla\Phi\right)^2 
                + \frac{\left[\sqrt{1+(\nabla h)^2}\partial_n\phi
                              +\nabla\Phi \cdot \nabla h\right]^2}
                       {2\left[1+(\nabla h)^2\right]}
\nonumber \\
                + g(t)h + \frac{\Gamma}{\rho}
                 \nabla\cdot\left(\frac{\nabla h}{\sqrt{1+(\nabla 
h)^2}}\right),
\label{eq:bcx2}
\eea
where $\partial_n\phi = \hat{n}\cdot(\nablatwo+\partial_z\hat{z})\phi$.  We 
note that except for the normal derivative $\partial_n\phi$, all other 
variables only depend on the two-dimensional coordinate $\vx$.
Since $\phi$ is a harmonic function, 
the normal derivative $\partial_n\phi$ at the boundary is related to the value
of $\phi$ at the boundary.  One such relation is given by the 
Dirichlet-Neumann 
operator $\sG(h)$.  The Dirichlet-Neumann operator $\sG(h)$ takes boundary 
values for a harmonic function and returns its normal derivative 
at the boundary with a metric pre-factor (Craig 1989; Craig \& Sulem 1993),
\be
\sG(h) \Phi(\vx,t) = \sqrt{1+(\nabla h)^2}\partial_n\phi.
\ee
We note that $\sG(h)$ is a linear operator for $\Phi(\vx,t)$ and
depends on the shape of the free surface $z=h(\vx,t)$ {\em nonlocally}.  

An important property of $\sG(h)$ is that it has a computable Taylor
expansion in powers of the surface displacement $h(\vx,t)$
and its spatial derivatives at $h(\vx,t) = 0$.  This Taylor expansion of
$\sG(h)$ is useful for studying the weakly nonlinear dynamics of surface
waves since only the first a few terms in the expansion, e.g., up to 
third order in $h(\vx,t)$, need to be evaluated.  The Taylor expansion 
of $\sG(h)$ up to order ${\cal O}(h^3)$ reads,
\bea
\sG(h)\Phi(\vx,t) = \sD\Phi - \nabla \cdot (h\nabla\Phi) -\sD(h\sD\Phi)
      + \sD\left[h\sD(h\sD\Phi) + \frac{1}{2}h^2\nabla^2\Phi\right]
\nonumber \\
+ \frac{1}{2} \nabla^2 (h^2\sD\Phi) 
-\nabla^2\left[\frac{1}{2}h^2\sD(h\sD\Phi) + \frac{1}{3}h^3\nabla^2\Phi\right]
\nonumber \\
\label{eq:DNoperator}
- \sD\left[h\sD(h\sD(h\sD\Phi)) + \frac{1}{2}h^2\nabla^2(h\sD\Phi)
                -\frac{1}{6}h^3\nabla^2(\sD\Phi)
                +\frac{1}{2} h\sD(h^2\nabla^2\Phi)\right],
\eea
where $\sD$ is a linear Fourier-integral operator and is defined for an
arbitrary function $u({\bf x})$ by 
$$
{\hat{{\cal D}}} u({\bf x}) =
\int_{-\infty}^{\infty}|{\bf k}|\hat{u}({\bf k}) \exp(i{\bf k}\cdot{\bf x})
d{\bf k}
$$ 
where $\hat{u}({\bf k})$ is the Fourier transform of $u({\bf x})$.
The operator $\sD$ is also nonlocal and is sometimes written as
$\sD  = \sqrt{-\nabla^2}$.
By substituting the expansion for the Dirichlet-Neumann operator $\sG(h)$
into the boundary conditions, and consistently keeping only terms up to the 
third order in $h$ and/or $\Phi$, we have
\bea
\label{eq:h1}
\partial_t h(\vx,t)  = 2\nu \nabla^2 h + \sD\Phi - \nabla \cdot (h\nabla\Phi)
             + \frac{1}{2} \nabla^2 (h^2\sD\Phi)
\nonumber \\
 -\sD(h\sD\Phi) + \sD\left[h\sD(h\sD\Phi) + \frac{1}{2}h^2\nabla^2\Phi\right],
\\
\label{eq:Phi1}
\partial_t \Phi(\vx,t) = 2\nu \nabla^2\Phi + g(t)h 
                + \frac{\Gamma}{\rho}\nabla^2 h
                + \frac{1}{2}\left(\sD\Phi\right)^2
                - \frac{1}{2}\left(\nabla\Phi\right)^2
\nonumber \\
     -(\sD\Phi)\left[h\nabla^2\Phi + \sD(h\sD\Phi)\right]
     -\frac{\Gamma}{2\rho}\nabla\cdot\left(\nabla h(\nabla h)^2\right).
\eea
Consideration of higher order terms is certainly not a difficulty in this
formulation, but most of the phenomena in Faraday waves near onset should 
be explained within a framework that includes up to third order 
nonlinearities.  
Equations (\ref{eq:h1}) and (\ref{eq:Phi1}) are two-dimensional, and they 
are the starting point for the analytical asymptotic analysis presented
below, and for extensive numerical studies that will be reported elsewhere
(a short summary of both analytical and numerical results can be found
in \cite{re:zhang96}).

Finally, we note that the incompressibility condition (Eq.~\ref{eq:incomp}) 
implies that the average level of the surface displacement $h(\vx,t)$ is  
constant, 
\be
\label{eq:average_h_constant}
\int_S h(\vx,t) d\vx = \mbox{constant}.
\ee
It is easy to see that $\int_S h(\vx,t) d\vx$ is indeed a constant of 
motion for Eq.~(\ref{eq:h1}).  

Before we proceed any further, it is useful to discuss at this point
the dissipation function approach used by other authors to obtain
dissipative contributions to the equation of motion 
for weakly damped waves, and compare it with the quasi-potential 
approximation. 
As we show below, both methods already differ in the linear viscous
terms in the dynamical equations even though, by construction,
they give the same (correct) rate of decay of the energy for {\em linear 
surface
waves}.  As a consequence, is seems to us that nonlinear viscous terms 
obtained 
from an energy balance on the inviscid amplitude equation may not be reliable.
 
It is well known that the irrotational surface wave problem in an inviscid 
fluid can be written in a Hamiltonian form (Zakharov 1968; Miles 1977).
The governing equations for irrotational surface waves can be written in this 
case as,
\bea
\label{eq:Hamiltonian_h}
\partial_{t}h(\vx,t) = \frac{\delta H}{\delta \Phi(\vx,t)},
\\
\label{eq:Hamiltonian_Phi}
\partial_{t}\Phi(\vx,t) = - \frac{\delta H}{\delta h(\vx,t)},
\eea
where $h(\vx,t)$ and $\Phi(\vx,t)$ are the generalized coordinate and momentum
respectively, and the Hamiltonian $H$ is given by
\be
\label{eq:hamiltonian_explicit}
H=\int\!\!\int d\vx \left[\frac{1}{2}\int^{h(\vx,t)}_{-\infty} dz
\biggl((\nablatwo\phi)^{2}+(\partial_z\phi)^2\biggr) + \frac{1}{2} g(t)h^{2} 
+ \Gamma\left(\sqrt{1\!+\!(\nabla h)^{2}}-1\right)\right].
\ee

This Hamiltonian formulation (or equivalently the corresponding Lagrangian 
formulation) offers a natural way to incorporate the effects of viscous
damping by adding a dissipation function. This
is done by modifying the equation for the generalized momentum $\Phi(\vx,t)$,
\be
\label{eq:gen_mom}
\partial_{t}\Phi(\vx,t) = - \frac{\delta H}{\delta h(\vx,t)} 
                        + Q(h(\vx,t), \Phi(\vx,t)),
\ee
where $Q(h, \Phi)$ is a dissipative or damping force, not invariant under
time reversal, and often of phenomenological nature. In the case of
Faraday waves,
the dissipation function $Q(h, \Phi)$ has been determined by
equating the rate of energy loss in this near-Hamiltonian formulation,
\be
\label{eq:dH}
\frac{dH}{dt}-\frac{\partial H}{\partial t} = \int\!\!\int d\vx \/
Q(h,\Phi) \partial_{t}h,
\ee
to the decay rate of the total energy for {\em potential flow} (Landau \&
Lifshitz 1959),
\be
\int\!\!\int d\vx \/ Q(h,\Phi) \partial_{t}h 
= -\nu\int\!\!\int d\vx \int_{-\infty}^{h(\vx,t)}
   \nabla^2 \left(\nabla \phi\right)^2.
\ee
It is easy to show that the linear part of $Q$ depends only on $\Phi(\vx,t)$,
and is given by
\be
\label{eq:diss_linear}
Q(h,\Phi) = 4\nu\nabla^2 \Phi(\vx,t) + \mbox{nonlinear terms}.
\ee
If we consider the linear approximation for $Q$, Eq.~(\ref{eq:Hamiltonian_Phi})
is modified by the addition of a viscous damping term $4\nu\nabla^2
\Phi(\vx,t)$ to the RHS, while Eq.~(\ref{eq:Hamiltonian_h}) remains unchanged. 
  
However, in the quasi-potential approximation (QPA), viscous damping terms
appear in both the $\partial_t h$ equation and the $\partial_t \Phi$ equation
(Eqs.~(\ref{eq:h1}) and (\ref{eq:Phi1})).  This difference has important
implications for the standing wave amplitude equations to be derived 
in~\S\,\ref{sec:sinusoidal}.  Recall that the viscous correction term in the 
$\partial_t h$ equation is related to the rotational component of the velocity
field at the fluid surface, while the viscous correction term in the 
$\partial_t
\Phi$ equation is related to the normal stress of the irrotational component.

For linear surface waves, both approaches give the same dynamical equation 
for a Fourier mode $\hat{h}_k(t)$ of $h(\vx,t)$,
\be
\partial_{tt} \hat{h}_k(t) + 4\nu k^2 \partial_{t}\hat{h}_k(t)
+\left(\frac{\Gamma k^3}{\rho} + g_0k + kg_z(t)\right)\hat{h}_k(t) = 0.
\ee
Thus, the decay rate of the total energy for linear surface
waves is the same for the two approaches, as expected.  

Since the calculation of the dissipation function $Q$ can be carried out for 
the nonlinear terms as well,  it would appear that the dissipative 
function approach is a natural way to incorporate nonlinear viscous terms order
by order. However, since the dissipation function approach does 
not give the correct linear viscous terms (Eqs. (\ref{eq:gen_mom}) and 
(\ref{eq:diss_linear}) lead to a linear viscous term in the equation for 
$\partial_{t} \Phi$ equal to $4 \nu \nabla^{2} \Phi$, whereas the linear viscous
term in Eq. (\ref{eq:Phi1}) is only $2 \nu \nabla^{2} \Phi$),
there does not seem to be any a priori reason
to trust nonlinear viscous terms.  

\section{Weakly nonlinear analysis}
\label{sec:sinusoidal}

Standing wave patterns of square symmetry are observed near onset in Faraday
experiments of weakly viscous fluids, in containers of large lateral size, 
and with single frequency sinusoidal forcing 
(Faraday 1831; Rayleigh 1883; Lang 1962; Ezerskii, Korotin \& Rabinovich
1985; Tufillaro, Ramshankar \& Gollub 1989; Ciliberto, Douady \& Fauve 1991;
Christiansen, Alstr{\o~}m \& Levinsen 1992; Bosch \& van de Water 1993;
Edwards \& Fauve 1993, 1994; M\"uller 1993).
We derive next
a set of coupled standing wave amplitude equations valid near onset that can
accommodate patterns of arbitrary symmetry on a two-dimensional surface.  
The standing wave amplitude equations that we will obtain are 
of gradient form, and thus minimization of the resulting Lyapunov functional 
determines the symmetry of the most stable standing wave state.  
Our derivation of the standing amplitude equations has three novel features.
The first one is the different starting point for the asymptotic expansion.
It is based on the LDQPE's described above. Second, we note that
there are two independent small parameters in this system, 
namely the reduced dimensionless driving amplitude $\varepsilon$, which is 
also the distance away from threshold, and the viscous damping parameter 
$\gamma$ (to be defined below).  
A double perturbative expansion for these two small parameters 
is necessary.  Solutions of the linearized quasi-potential equations are
obtained by performing a perturbative expansion for the small damping 
parameter or the driving amplitude $f$ (to be defined below).  
The linear solutions 
contain the primary mode of the fluid surface with a frequency half of 
the driving frequency as well as its higher harmonics.  These higher harmonic 
terms are proportional to the driving amplitude $f$ or its powers ($f^n$, with
$n=2,3 \cdots.$). The nonlinear interaction of these higher harmonic terms 
with the primary mode provides a novel {\em amplitude-limiting effect} for the 
parametric surface wave system.  This effect is important for the nonlinear 
saturation of the surface wave amplitude in weakly dissipative systems.  

The third feature is related to three-wave resonant interactions in
capillary-gravity surface waves.  Although quadratic terms are prohibited 
by symmetry in the standing wave amplitude equations that we derive, 
three-wave resonance
(triad resonance) plays a crucial role in pattern selection.
Both three- and four-wave resonant interactions 
among capillary-gravity waves are well known and well studied.  The importance
of three-wave resonant interactions to pattern selection in Faraday waves,
however, has been largely overlooked.  As we show later, 
the resonant interactions between two linearly unstable standing wave modes 
and a linearly stable wave mode strongly affects four-wave nonlinear 
interactions, and thus the coefficient of third order nonlinear terms in 
the amplitude equations.

Previous theoretical work by \cite{re:milner91} involved the
derivation of a set of coupled {\em traveling} wave 
amplitude equations for inviscid parametric surface waves, to which viscous 
damping terms were added by an energy-balance-consideration, equivalent
to the dissipation function approach describe above.  He concluded that 
nonlinear viscous 
damping terms in the dissipation function play a major role in
pattern selection. We disagree with his conclusion
for four reasons: 
(i) As discussed earlier, the
dissipation function approach does not give the correct linear viscous terms,
so it is doubtful that it will introduce the correct nonlinear damping terms;
(ii) linear viscous terms in the fluid 
equations can contribute to nonlinear damping terms in the amplitude 
equations, while such contribution is absent in Milner's phenomenological 
consideration of viscous effects; (iii) an amplitude-limiting effect of 
the driving force did not appear in Milner's analysis since he used 
a zeroth order linear solution for the parametric instability; and (iv) Milner 
did obtain triad resonant interactions in his calculation, but by not taking 
them
into account explicitly, he overlooked their effect on pattern selection.
As was recently suggested by \cite{re:edwards94}, 
we will show that triad resonant interactions play an important role in 
pattern formation of Faraday waves in weakly viscous fluids.

\subsection{Solutions of the Linearized Equations}
\label{sec:sinusoidal-linear}

As is well known,  the linearized problem of parametric 
surface waves can be reduced to the damped Mathieu equation, and
the Faraday instability corresponds to the subharmonic resonance of the 
equation.
For the case of a sinusoidal driving force, the effective acceleration $g(t)$
in Eqs.~(\ref{eq:h1}) and (\ref{eq:Phi1}) can be written as
$ g(t) = -g_0 - g_{z} \sin \Omega t,$
where $g_0$ is the constant acceleration of gravity, and $\Omega$ and $g_{z}$ 
are the angular frequency and the amplitude of the driving force respectively. 
We now choose $2/\Omega \equiv 1/\omega_0$ as the unit of time and $1/k_0$ as 
the unit of length with $k_0$ defined by $ 
\omega_0^2 = g_0 k_0 + \frac{\Gamma}{\rho}k_0^3.$ 
We also choose the unit for the surface velocity potential $\Phi$ to be
$\omega_0/k_0$. We further define a
dimensionless linear damping coefficient $\gamma = 
2\nu k_0^2/\omega_0$, $G_0 = g_0k_0/\omega_0^2$, 
$\Gamma_0 = \Gamma k_0^3/(\rho\omega_0^2)$, and the dimensionless driving
amplitude $f = g_{z}k_0/(4\omega^2_0)$.  Note that $G_0 + \Gamma_0 = 1$ by 
definition. 

By linearizing the quasi-potential equations and boundary conditions
(Eqs.~(\ref{eq:h1}) and (\ref{eq:Phi1}))
with respect to the surface displacement $h$, and the surface
velocity potential $\Phi$, and taking the Fourier transform with respect to the
spatial coordinate, one obtains in a standard way,
\bea
\label{eq:linear_h2}
\partial_{tt}\hat{h}_k + 2\gamma k^2 \partial_t \hat{h}_k 
+\left(G_0 k+\Gamma_0 k^3 + \gamma^2 k^4 + 4fk \sin 2t\right) \hat{h}_k = 0, \\
\label{eq:linear_Phi2}
k \hat{\Phi}_k = \partial_t \hat{h}_k + \gamma k^2 \hat{h}_k.
\eea
Equation (\ref{eq:linear_h2}) is the damped Mathieu equation for $\hat{h}_k$.
We now seek analytical solutions of the above equations perturbatively.
We introduce a small parameter $\eta$ ($\eta \ll 1$) such that
$\mbox{$\gamma = \gamma_0 \eta$, and $f = f_0 \eta$},$ 
where $\gamma_0$ and $f_0$ are assumed to be of ${\cal O}(1)$.  
For weakly dissipative fluids, i.e. $\gamma \ll 1$, and to be consistent
with the quasi-potential approximation discussed in last section,
we will neglect the term proportional to $\gamma^2$ in Eq.~(\ref{eq:linear_h2}).
At subharmonic resonance, we have $\omega^{2}(k)
\equiv G_0 k+\Gamma_0 k^3 = 1$, which of course implies $k = 1$ (note that
$G_0+\Gamma_0=1$ by definition).  
We then consider an expansion for the wavenumber $k$ near subharmonic
resonance as, $ k = 1 + \Delta k \eta + \cdots.$
Above (and near) the onset of subharmonic resonance, we expect the amplitudes
of $\hat{h}_k$ to grow in time but in a slower time scale than that for 
the subharmonic oscillation.  In the following we assume that the slow time is 
$T = \eta t$, and seek solutions perturbatively as power series in $\eta$, 
\bea
\hat{h}_k = \hat{h}_k^{(0)}(t, T) + \eta \hat{h}_k^{(1)}(t, T) + \cdots,
\nonumber \\
\hat{\Phi}_k = \hat{\Phi}_k^{(0)}(t, T) + \eta \hat{\Phi}_k^{(1)}(t, T) + 
\cdots. 
\nonumber
\eea
At  ${\cal O}(\eta^0)$, one has,
\bea
\hat{h}_k^{(0)}(t, T) = A_k(T) \cos t + B_k(T) \sin t, 
\nonumber \\
\hat{\Phi}_k^{(0)}(t, T) = -A_k(T)\sin t + B_k(T) \cos t,
\nonumber
\eea
where $A_k(T)$ and $B_k(T)$ are arbitrary functions.
At ${\cal O}(\eta^1)$, a standard solvability condition appears,
\bea
\label{eq:solvability_linear_A}
\partial_TA = (f_0-\gamma_0)A + \frac{1}{2}(G_0+3\Gamma_0)\Delta k B, \\
\label{eq:solvability_linear_B}
\partial_TB = -(f_0+\gamma_0)B - \frac{1}{2}(G_0+3\Gamma_0)\Delta k A.
\eea
By substituting $A$ and $B \propto e^{\sigma t}$, one finds
\begin{equation}
\label{eq:growth_rate}
\sigma_{\pm} = - \gamma_{0} \pm \sqrt{f_{0}^{2} - \left( \frac{\Delta k}{2}
(G_{0}+ 3 \Gamma_{0} )\right)^{2}}.
\end{equation}
Exactly at subharmonic resonance ($k=1$ or $\Delta k = 0$), the 
growing mode $M_+ \propto A$ and the decaying mode $M_- \propto B$.  In 
this case, the linearly growing eigenmode above onset
is given by,
\bea
\label{eq:linear_h_x2}
h(\vx, t) = \left(\cos t + \frac{f}{4} \sin 3t + \cdots \right)
            \sum_{j=1}^{N}\left[A_j(t) \exp \left(i\hkj\cdot\vx\right)
                                + c.c.\right], \\
\label{eq:linear_Phi_x2}
\Phi(\vx, t) = \left(-\sin t + f \cos t + \frac{3f}{4}\cos 3t + \cdots \right)
             \sum_{j=1}^{N}\left[A_j(t) \exp \left(i\hkj\cdot\vx\right)
                                + c.c.\right],
\eea
where we have assumed that the standing waves consist of an arbitrary 
discrete set of wavevectors in the two-dimensional space.  When $f=\gamma$ 
(at onset), Eqs.~(\ref{eq:linear_h_x2}) and (\ref{eq:linear_Phi_x2}) are the 
linear neutral solutions, which is the basis of a weakly nonlinear analysis 
for our problem.  It is important to note at this point that we have kept in 
the linear solution terms proportional 
to $f$ (or $\gamma$ since $f=\gamma$ at onset). These terms
will be 
crucial for obtaining the correct cubic term in the amplitude equations, 
and had not been included in previous studies.
Terms proportional to higher 
harmonics do not contribute to the standing wave amplitude equations to the
order considered.

\subsection{Standing Wave Amplitude Equations}
\label{sec:SWAE}

We seek nonlinear standing wave solutions of Faraday waves near 
onset ($\varepsilon \equiv (f-\gamma)/\gamma \ll 1$) in this section.  
We expand the two-dimensional 
quasi-potential equations (Eqs.~(\ref{eq:h1}) and (\ref{eq:Phi1})) 
consistently in $\varepsilon^{1/2}$ with multiple time scales,
\bea
\label{eq:expansion_h}
h(\vx, t, T) = \varepsilon^{1/2} h_1(\vx, t, T) + \varepsilon h_2 
             + \varepsilon^{3/2} h_3 + \cdots, \\
\label{eq:expansion_Phi}
\Phi(\vx, t, T) = \varepsilon^{1/2} \Phi_1(\vx, t, T) + \varepsilon \Phi_2
             + \varepsilon^{3/2} \Phi_3 + \cdots,
\eea
where the slow time scale $T=\varepsilon t$, and is not related to the slow 
time
in the previous section. The scaling of the amplitudes and the slow time $T$ 
with
$\varepsilon$ is formally determined by the ultimate consistency of the
expansion, and in particular by the balance of terms in 
the final form of the standing wave amplitude equation (Eq.~(\ref{eq:SWA1})).
We define the following linear operator,
\be
{\cal L} \equiv \left(\begin{array}{cc}
                 \partial_t - \gamma\nabla^2  & -\sD \\
                 G_0-\Gamma_0\nabla^2+4\gamma\sin 2t & 
\partial_t-\gamma\nabla^2
            \end{array}
      \right).
\ee
On substituting Eqs.~(\ref{eq:expansion_h}) and (\ref{eq:expansion_Phi}) into 
the quasi-potential equations we have at ${\cal O}(\varepsilon^{1/2})$,
$$ 
{\cal L}\left(\begin{array}{c}
              h_1 \\
              \Phi_1
              \end{array}
        \right) = 0.
$$ 
The above equation is the same as the linearized quasi-potential equations
discussed in the last section except the driving amplitude $f$ is replaced by
the damping coefficient $\gamma$.  Thus, the solutions for $h_1$ and $\Phi_1$ 
are just the linear solutions found in the last section.  For simplicity, we 
will neglect the linearly stable mode $B$ of the linear solutions.  Then, 
$h_1$ and $\Phi_1$ are the linear neutral solutions, 
\bea
\label{eq:order1_h}
h_1 = \left(\cos t + \frac{\gamma}{4} \sin 3t\right)
                 \sum_{j=1}^{N}\left[A_j(T) \exp \left(i\hkj\cdot\vx\right)
                                + c.c.\right], \\
\label{eq:order1_Phi}
\Phi_1 = \left(-\sin t + \gamma\cos t + \frac{3\gamma}{4}\cos 3t\right)
                  \sum_{j=1}^{N}\left[A_j(T) \exp \left(i\hkj\cdot\vx\right)
                                + c.c.\right].
\eea

At ${\cal O}(\varepsilon)$, we have
\be
\label{eq:order2}
{\cal L}\left(\begin{array}{c}
              h_2 \\
              \Phi_2
              \end{array}
        \right) = \left(\begin{array}{c}
                         -\nabla\cdot\left(h_1\nabla\Phi_1\right)
                         -\sD\left(h_1\sD\Phi_1\right) \\
                         \frac{1}{2}(\sD\Phi_1)^2 - \frac{1}{2}(\nabla\Phi_1)^2
                         \end{array}
                  \right).
\ee
The above equation represents two coupled equations for $h_2$ and $\Phi_2$.
It is easy to obtain an independent equation for $h_2$ from
Eq.~(\ref{eq:order2}), which reads,
\bea
\partial_{tt}h_2 - 2\gamma\nabla^2\partial_t h_2 
+ (G_0-\Gamma_0\nabla^2)\sD h_2 + 4\gamma h_{2} \sin 2t 
\nonumber \\
= \frac{1}{2}\sD\left[(\sD\Phi_1)^2-(\nabla\Phi_1)^2\right]  
+\left(\gamma\nabla^2-\partial_t\right)
 \left[\nabla\cdot\left(h_1\nabla\Phi_1\right) 
       +\sD\left(h_1\sD\Phi_1\right)\right] 
\nonumber \\
= \sum_{j,l=1}^{N}\Biggl\{\frac{1+c_{jl}}{4}\sqrt{2(1+c_{jl})}
  -\cos 2t \left[1+c_{jl}-\frac{3-c_{jl}}{4}\sqrt{2(1+c_{jl})}\right]
\nonumber \\
  -\gamma\sin 2t \left[(5/2+c_{jl})\left(1+c_{jl}-\sqrt{2(1+c_{jl})}\right)
  +\frac{1+c_{jl}}{8}\sqrt{2(1+c_{jl})}\right]\Biggr\}
\nonumber \\
  \left[A_jA_l\exp\left(i(\hkj+\hkl)\cdot\vx\right) + c.c.\right] 
\nonumber \\
+ \sum_{j,l=1}^{N}\Biggl\{\frac{1-c_{jl}}{4}\sqrt{2(1-c_{jl})}
  -\cos 2t \left[1-c_{jl}-\frac{3+c_{jl}}{4}\sqrt{2(1-c_{jl})}\right]
\nonumber \\
  -\gamma\sin 2t \left[(5/2-c_{jl})\left(1-c_{jl}-\sqrt{2(1-c_{jl})}\right)
  +\frac{1-c_{jl}}{8}\sqrt{2(1-c_{jl})}\right]\Biggr\}
\nonumber \\
\label{eq:h_order2}
  \left[A_jA_l^*\exp\left(i(\hkj-\hkl)\cdot\vx\right) + c.c.\right] + \cdots,
\eea
where $c_{jl} \equiv \cos \theta_{jl} = \hkj\cdot\hkl $ and we have 
neglected terms that are proportional to $\gamma^2$.  We note that 
there are no terms on the RHS proportional to $\cos t$ or $\sin t$, that
would introduce a secular variation in the solution.  Therefore, there is 
no solvability condition for the amplitudes $A_j$ at this order.  There are, 
however, resonant interactions due to certain terms on the RHS.

The particular solution $h_2$ of Eq.~(\ref{eq:h_order2})  can be written as
\bea
h_2 = \sum_{j,l=1}^{N}\biggl\{
    H_{jl}(t)\left[A_jA_l\exp\left(i(\hkj+\hkl)\cdot\vx\right) + c.c.\right]
\nonumber \\
   +H_{j,-l}(t)\left[A_jA_l^*\exp\left(i(\hkj-\hkl)\cdot\vx\right)+ c.c.\right],
\eea
where $H_{jl}(t)$ is an unknown function to be determined, and $H_{j,-l}$ is
defined by replacing $c_{jl}$ with $c_{j,-l}$ in $H_{jl}$.  On substituting
the above form for $h_2$ into Eq.~(\ref{eq:h_order2}), we have the following
equation for $H_{jl}(t)$,
\bea
\partial_{tt}H_{jl} + 2\gamma\sqrt{2(1+c_{jl})}\partial_t H_{jl}
+ [G_0+2\Gamma_0(1+c_{jl})]\sqrt{2(1+c_{jl})}H_{jl} 
\nonumber \\
\label{eq:resonance_explained}
= F_{jl}^{(1)}\cos 2t + F_{jl}^{(2)}\sin 2t + \cdots,
\eea
where $F_{jl}^{(1)}$ and $F_{jl}^{(2)}$ are proportional to $A_jA_l$.
Only terms that are relevant to the resonance are written out on the RHS of 
Eq.~(\ref{eq:resonance_explained}).  Equation (\ref{eq:resonance_explained}) 
looks very much like the equation for an additively forced harmonic oscillator 
with friction.  When the ``natural" frequency of the ``oscillator", 
$[(G_0+2\Gamma_0(1+c_{jl}))\sqrt{2(1+c_{jl})}]^{1/2}$, equals the driving 
frequency, resonance occurs.  This condition reads,
\be
\label{eq:triad_resonance_condition}
[G_0+2\Gamma_0(1+c_{jl})]\sqrt{2(1+c_{jl})} = 4.
\ee
Due to the nonzero damping coefficient, $2\gamma\sqrt{2(1+c_{jl})}$, this 
resonance results in a finite value for $H_{jl}$ that is inversely proportional
to the damping coefficient.  We note that the parametric forcing term 
$h_{2} \sin 2t$ in Eq.~(\ref{eq:h_order2}) is not directly relevant to the 
resonant interaction.

\begin{figure}
\vspace{3in}
\includegraphics{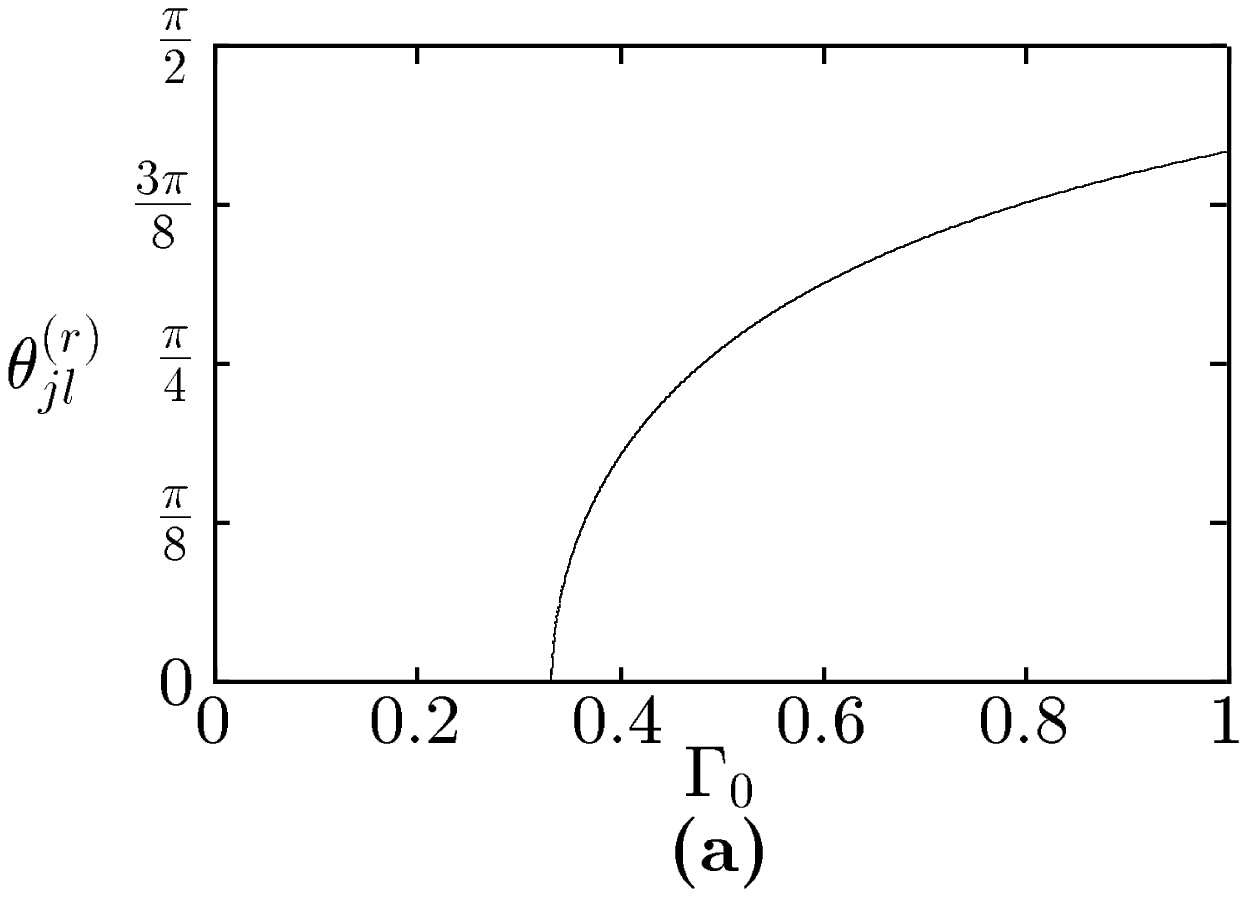}
\includegraphics{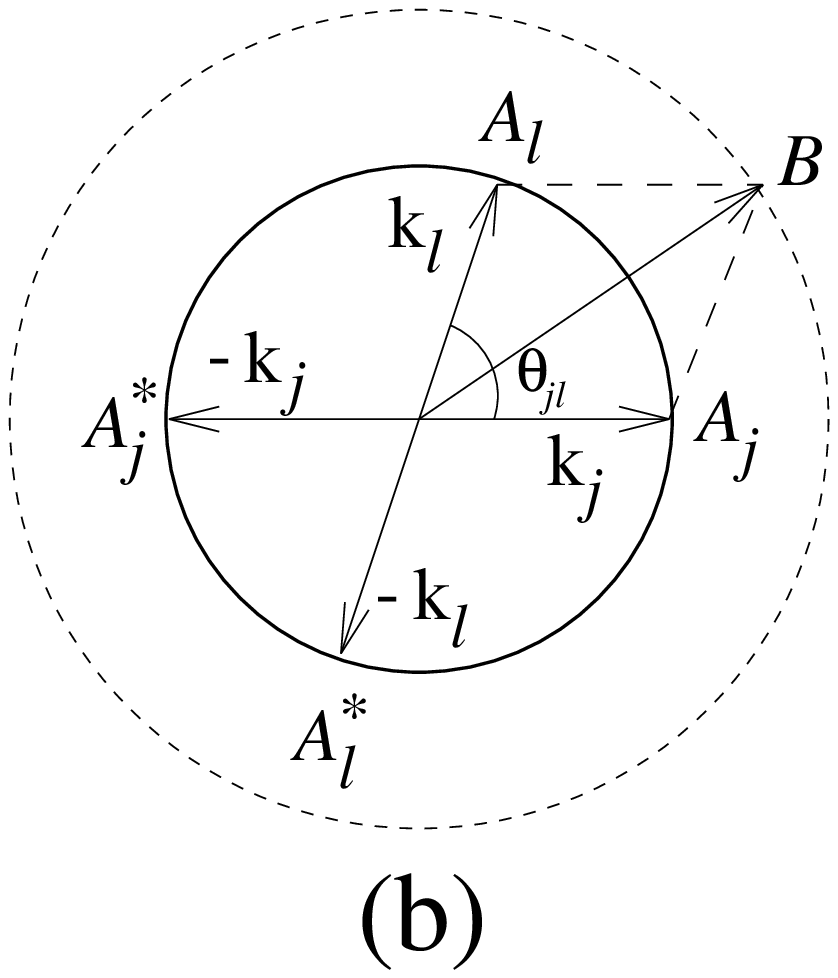}
\caption{Triad resonant interaction in parametric surface waves (a)
$\theta_{jl}^{(r)}$ as a function of $\Gamma_0$.  (b) The mode $B$ will
resonate with the quadratic interaction of the standing waves $A_j$
and $A_l$ when $\theta_{jl} = \theta_{jl}^{(r)}$.}
\label{fig-triad}
\end{figure}

The values of $\theta_{jl}$ ($c_{jl} = \cos\theta_{jl}$) that satisfy the 
resonance condition (Eq.~(\ref{eq:triad_resonance_condition})) as a 
function of $\Gamma_0$ are shown in Fig.~\ref{fig-triad}(a).
Since the RHS of Eq.~(\ref{eq:resonance_explained}) is proportional to 
$A_jA_l$, there are three waves involved in this resonance, namely,
standing wave modes $A_j$ and $A_l$, and mode $B$ with wavevector
$\hkj+\hkl$, as shown in Fig.~\ref{fig-triad}(b).   Therefore, 
Eq.~(\ref{eq:resonance_explained}) describes a three-wave resonant interaction.
Note that the wavenumber for mode $B$ is away from the critical wavenumber
$k_0=1$, and thus mode $B$ is a linearly stable mode.  
For $\Gamma_0 < 1/3$, triad resonance is not possible. 
For $\Gamma_0=1/3$, wavevectors of the three resonating waves are in the 
same direction ($\theta_{jl}^{(r)} = 0$).  
As $\Gamma_0$ is further increased, 
$\theta_{jl}^{(r)}$ also increases.  For purely capillary waves ($\Gamma_0 = 
1$), $\theta_{jl}^{(r)}$ reaches the maximum value of $c_{jl} = 2^{1/3}-1$ or 
$\theta_{jl}^{(r)} \approx 74.9^{\circ}$.  

Resonant interactions among surface capillary-gravity waves in general have 
been studied since the pioneering work by \cite{re:wilton15} (for
a recent review see Hammack \& Henderson 1993).
Wilton found that a Stokes expansion 
in powers of the wave amplitude became singular at a wavenumber 
$k=\sqrt{g_0\rho/2\Gamma}$ for inviscid capillary-gravity 
waves in two spatial dimensions, 
which corresponds to the triad resonance discussed above for $\Gamma_0 = 1/3$. 
 
This special case of resonant interaction is often termed {\em second-harmonic 
resonance\/} and the corresponding capillary-gravity waves are called 
Wilton's ripples.  Experimental studies by \cite{re:mcgoldrick70}, 
and \cite{re:banerjee82} on the triad resonance in 
capillary-gravity waves have verified the function $\theta_{jl}^{(r)}$ 
(see Fig.~\ref{fig-triad}(a)) quantitatively (Hogan 1984).  

The relevance of triad resonant interactions at second order to pattern
formation in Faraday waves has been largely overlooked by previous studies. 
We want to emphasize here that the effect of triad
resonant interactions on pattern formation in parametric surface waves can be 
intuitively understood.  Let us consider a situation in which two linearly 
unstable 
standing waves $A_j$ and $A_l$ with their wavevectors separated exactly the 
resonating angle $\theta_{jl}^{(r)}$ grow from small amplitudes in the linear 
regime.  When they enter the nonlinear regime, a mode $B$ with wavevector 
$\hkj+\hkl$ is created as a results of quadratic nonlinear interaction of $A_j$
and $A_l$.  The amplitude of $B$ will become very large because of the 
resonance 
and weak damping.  For inviscid fluids, the amplitude of $B$ will increase 
without limit.  Since the parametric forcing pumps energy into the surface wave
system through unstable modes $A_j$ and $A_l$ at a rate determined by the
supercriticality $\varepsilon$, the energy required for the growth of mode 
$B$ will come from reductions of the amplitudes in $A_j$ and $A_l$, but not
directly from the external parametric force.  As a result, the modes $A_j$ and 
$A_l$ will have smaller amplitudes than other unstable modes that do not 
satisfy the triad resonance condition.  In other words,
small perturbations with components close to the critical wavenumber will 
grow exponentially.
When the amplitudes of these modes become large enough, a nonlinear selection
process takes place such that modes with their wavevectors
separated by the resonating angle $\theta_{jl}^{(r)}$ are less favored or 
avoided.
The above argument is relevant for both inviscid or weakly viscous fluids.  
For fluids of high viscosity, the influence of triad resonant interactions
becomes smaller because of the large damping.

Since there is no solvability condition at second order, there will be no 
quadratic terms in the amplitude equations to be derived.   Thus, the above 
described influence of triad resonant interactions must appear through cubic 
nonlinear terms, which represents four-wave resonant 
interactions among four linearly unstable standing wave modes.  
Specifically, in particular
the coefficient of the cubic nonlinear terms in the amplitude equations are
expected to have a large positive peak at the resonating angle 
$\theta_{jl}^{(r)}$ as we show later in this section.
When viscous damping
effects are neglected, the peak shifts toward $+\infty$.  Such divergence of 
the nonlinear interaction coefficient was encountered by \cite{re:milner91} 
in his analysis of weakly damped parametric surface waves because
he neglected the contribution to the amplitude equation from the second
order solution. As a consequence, he failed to realize the relevance of such 
a divergence to pattern selection. In contrast, and in agreement with our
calculations,  \cite{re:edwards94} had suggested 
recently that triad resonance could be important for pattern selection.

As we show later, the occurrence of standing wave pattern of 
square symmetry in capillary Faraday waves is closely related to triad 
resonance
with $\theta_{jl}^{(r)}=74.9^{\circ}$.  By increasing the gravity wave 
component 
in capillary-gravity waves, the resonating angle $\theta_{jl}^{(r)}$ becomes 
smaller and the nonlinear interaction coefficients at cubic order changes 
accordingly.  For $\Gamma_0 \approx 1/3$, $\theta_{jl}^{(r)}$ is close to zero 
and we will show that standing wave patterns of square symmetry become 
unstable 
and hexagonal, triangular, or quasicrystalline patterns can be stabilized. 

The solution of the corresponding homogeneous equation of 
Eq.~(\ref{eq:h_order2}) will have the same form as the solution at order
${\cal O}(\varepsilon^{1/2})$, and thus can be absorbed into the solution
at ${\cal O}(\varepsilon^{1/2})$.  As a result, we are only interested 
in the particular solutions to the inhomogeneous equation for $h_2$.  
The particular solutions are,
\bea
h_2 = \sum_{j,l=1}^{N}\biggl\{
    \left(\alpha_{jl}+\beta_{jl}\cos 2t+\gamma\delta_{jl}\sin 2t\right)
    \left[A_jA_l\exp\left(i(\hkj+\hkl)\cdot\vx\right) + c.c.\right]
\nonumber \\
\label{eq:h_order2_SWAE}
   +\left(\bar{\alpha}_{jl}+\bar{\beta}_{jl} \cos 2t + 
     \gamma\bar{\delta}_{jl} \sin 2t\right)
    \left[A_jA_l^*\exp\left(i(\hkj-\hkl)\cdot\vx\right) + c.c.\right]\biggr\},
\eea
where
\bea
\alpha_{jl} = \frac{1+c_{jl}}{4\left[G_0+2\Gamma_0(1+c_{jl})\right]}
            - \frac{2\gamma^2 \delta_{jl}}{G_0 + 2\Gamma_0 (1+c_{jl})}, 
\\
\beta_{jl}  = 
\frac{-\left(1+c_{jl}-\frac{3-c_{jl}}{4}\sqrt{2(1+c_{jl})}\right)
       \left(D_{jl} - 8\gamma^2 M_{jl}\right) + 8\gamma^2(1+c_{jl}) N_{jl}}
     {64 \gamma^2(1+c_{jl})^2 + D_{jl}^2 - 8\gamma^2 D_{jl} M_{jl}},
\\
\delta_{jl} = 
-\frac{8(1+c_{jl})\left(1+c_{jl}-\frac{3-c_{jl}}{4}\sqrt{2(1+c_{jl})}\right)
      + D_{jl}N_{jl}}
     {64 \gamma^2(1+c_{jl})^2 + D_{jl}^2 - 8\gamma^2 D_{jl} M_{jl}},
\eea
and  
\bea
\label{eq:M_jl}
M_{jl} = \frac{\sqrt{2(1+c_{jl})}}{G_0+2\Gamma_0(1+c_{jl})}, \\
\label{eq:D_jl}
D_{jl} = \left[G_0+2\Gamma_0(1+c_{jl})\right]\sqrt{2(1+c_{jl})} - 4, \\
\label{eq:N_jl}
N_{jl} = \left(\frac{5}{2}+c_{jl}\right)\left(1+c_{jl}-\sqrt{2(1+c_{jl})}\right
)
       + \frac{1+c_{jl}}{8}\sqrt{2(1+c_{jl})} + (1+c_{jl})M_{jl}.
\eea
$\bar{\alpha}_{jl}$, $\bar{\beta}_{jl}$, and $\bar{\delta}_{jl}$ can be
obtained by replacing $c_{jl}$ with $-c_{jl}$ in the expressions for
$\alpha_{jl}$, $\beta_{jl}$, and $\delta_{jl}$ respectively.  The 
triad resonance occurs when $D_{jl}=0$.  We have retained terms that 
are proportional to $\gamma^2$ in the expressions for $\alpha_{jl}$, 
$\beta_{jl}$, and $\delta_{jl}$ since these terms become important 
when $D_{jl}$ is small or zero, i.e., at triad resonance.  The factor 
$\gamma^2$ in these terms will either cancel to give terms of ${\cal O}(1)$,
or partially cancel to give terms of ${\cal O}(1/\gamma)$.  Although we still
keep them when away from resonance, these terms are very small for weak 
damping, 
and thus should not affect the consistency of the perturbation expansion. 

From Eq.~(\ref{eq:order2}), we have
\be
\label{eq:Phi_2fromh_2}
\sD\Phi_2 = \partial_t h_2 - \gamma \nabla^2 h_2 
          + \nabla \cdot \left(h_1\nabla\Phi_1\right)
          + \sD\left(h_1\sD\Phi_1\right).
\ee
On substituting the expressions for $h_1$, $\Phi_1$, and $h_2$ into
the above equation, we obtain the following expression for $\Phi_2$,
\bea
\Phi_2 = \sum_{j,l=1}^{N}\biggl\{
    \left(\gamma u_{jl}+\gamma v_{jl}\cos 2t + w_{jl}\sin 2t\right)
    \left[A_jA_l\exp\left(i(\hkj+\hkl)\cdot\vx\right) + c.c.\right]
\nonumber \\
\label{eq:Phi_order2_SWAE}
   +\left(\gamma\bar{u}_{jl}+\gamma\bar{v}_{jl}\cos 2t+\bar{w}_{jl}\sin 
2t\right)
    \left[A_j A_l^*\exp\left(i(\hkj-\hkl)\cdot\vx\right) + c.c.\right]\biggr\},
\eea
where
\bea
u_{jl} = \frac{1}{2} + \left(\alpha_{jl}-\frac{1}{4}\right)\sqrt{2(1+c_{jl})},
\\
\label{eq:vjl}
v_{jl} = \frac{3}{4} + \left(\beta_{jl}-\frac{3}{8}\right)\sqrt{2(1+c_{jl})}
       + \frac{2\delta_{jl}}{\sqrt{2(1+c_{jl})}},
\\
\label{eq:wjl}
w_{jl} = -\frac{1}{2} + \frac{1}{4}\sqrt{2(1+c_{jl})} 
       - \frac{2\beta_{jl}}{\sqrt{2(1+c_{jl})}},
\eea
and $\bar{u}_{jl}$, $\bar{v}_{jl}$, and $\bar{w}_{jl}$ can be
obtained by replacing $c_{jl}$ with $-c_{jl}$ in the expressions for
$u_{jl}$, $v_{jl}$,
and $w_{jl}$ respectively.  We also note that $v_{jl}$ and $w_{jl}$ are
not singular at $c_{jl}=-1$ since the factor in denominator 
$\sqrt{2(1+c_{jl})}$ will 
be canceled by the same factor in the numerator in Eqs.~(\ref{eq:vjl}) and
(\ref{eq:wjl}).  Similarly, $\bar{v}_{jl}$ and $\bar{w}_{jl}$ are also not
singular at $c_{jl}=1$.

The amplitude $A_j$'s are not determined yet at ${\cal O}(\varepsilon)$, and 
thus it is necessary to continue the expansion to higher orders. 
At ${\cal O}(\varepsilon^{3/2})$, we have
\bea
{\cal L}\left(\begin{array}{c}
              h_3 \\
              \Phi_3
              \end{array}
        \right) = \left(\begin{array}{c}
                  -\partial_T h_1 
                  - \nabla\cdot\left(h_1\nabla \Phi_2 + h_2\nabla \Phi_1\right)
                  - \sD\left(h_1\sD\Phi_2 + h_2\sD\Phi_1\right) \\
                  - \partial_T \Phi_1 - 4\gamma h_{1} \sin 2t 
                  + \sD\Phi_1 \sD \Phi_2 - \nabla \Phi_1 \cdot \nabla \Phi_2
                       \end{array}
                  \right) 
\nonumber \\
\label{eq:third_order}
                + \left(\begin{array}{c}
                  \sD\left[h_1\sD(h_1\sD\Phi_1) 
                           + \frac{1}{2}h_1^2\nabla^2\Phi_1\right]
                  + \frac{1}{2}\nabla^2\left(h_1^2\sD\Phi_1\right) \\
                  -\sD\Phi_1 \left[h_1\nabla^2\Phi_1+\sD(h_1\sD\Phi_1)\right]
                  -\frac{\Gamma_0}{2}
                        \nabla\cdot\left(\nabla h_1 (\nabla h_1^2)\right)
                       \end{array}
                  \right).
\eea
We have found that it is convenient to obtain the solvability condition at this
order from the independent equation for $h_3$, which can be easily obtained 
from 
Eq.~(\ref{eq:third_order}), and it reads,
\bea
\partial_{tt}h_3 - 2\gamma\nabla^2\partial_t h_3 
+ (G_0\!-\!\Gamma_0\nabla^2)\sD h_3 + 4\gamma \sin 2t \sD h_3 
= -\partial_T \left((\gamma\!+\!\partial_t)h_1 \!+\! \Phi_1\right)
\nonumber \\
- 4\gamma h_{1} \sin 2t 
+ \left(\gamma\nabla^2-\partial_t\right)
  \biggl[\nabla\cdot\left(h_1\nabla \Phi_2 + h_2\nabla \Phi_1\right)
      +\sD\left(h_1\sD\Phi_2 + h_2\Phi_1\right)
\nonumber \\
      +\sD\left(\frac{1}{2}h_1^2\Phi_1-h_1\sD(h_1\Phi_1)\right)
      -\frac{1}{2}\nabla^2\left(h_1^2\Phi_1\right) \biggr]
\nonumber \\
\label{eq:third_order2}
+ \sD\left[\Phi_1\sD\Phi_2 \!-\! \nabla\Phi_1\cdot\!\nabla\!\Phi_2 + \Phi_1^2 
h_1
  -\Phi_1\sD(h_1\Phi_1) 
  - \frac{\Gamma_0}{2}\!\nabla\!\cdot\left(\nabla h_1 (\nabla h_1)^2\right) 
\right].
\eea
The Fredholm alternative theorem
requires that the RHS of Eq.~(\ref{eq:third_order2}) be orthogonal to 
any of the independent solution of its adjoint homogeneous equation.
The solvability condition reads,
\be
\label{eq:SWAE_solvability}
\int_0^{2\pi} dt \int_0^{2\pi} d\Theta_j  RHS_3 \tilde{h}_j(\vx,t) = 0,
\ee
where $RHS_3$ stands for the RHS of Eq.~(\ref{eq:third_order2}), and
$\tilde{h}_j(\vx, t)$ is either one of the following two independent solutions 
of the adjoint homogeneous equation, 
$$
\left(\cos t + \frac{\gamma}{4}\sin 3t + \cdots\right)
\exp\left(\pm i\hkj \cdot \vx\right),
$$
$$
\left(\sin t - \frac{\gamma}{4}\cos  3t + \cdots\right)
\exp\left(\pm i\hkj \cdot \vx\right),
$$
where $\hkj$ is a unit vector in arbitrary direction, and 
$\Theta_j=\hkj\cdot\vx$.  
Since both $RHS_3$ and $\tilde{h}_j(\vx,t)$ have a temporal period $2\pi$, we 
have 
taken the interval of integration in Eq.~(\ref{eq:SWAE_solvability}) to be 
$2\pi$.
Also because of the periodicity of $RHS_3$ and $\tilde{h}_j(\vx,t)$, it is 
only 
necessary to consider the term proportional to $\cos t$ or $\sin t$ in 
$\tilde{h}_j(\vx,t)$.  It turns out that the relevant solvability condition
is obtained from the second solution of $\tilde{h}_j(\vx,t)$.  If we had also 
considered the linearly stable mode $B_j$ in this analysis, another 
solvability 
condition would be obtained from the first solution of $\tilde{h}(t)$.
In summary, the solvability condition for Eq.~(\ref{eq:third_order2}) is that
the coefficient of $\sin t \exp\left(\pm i\hkj \cdot \vx\right)$ term in 
$RHS_3$ equals zero.

In what follows, we will collect terms from the RHS of 
Eq.~(\ref{eq:third_order2})
that are relevant to the solvability condition in a tedious but 
straightforward 
calculation.  These terms can be written as 
\bea
RHS = 2\sin t \sum_{j=1}^{N}\Biggl\{
 \Biggl[\partial_TA_j - \gamma A_j 
       +\gamma\Biggl(\left( \frac{28+9\Gamma_0}{64}
                           +2\alpha_{jj}+\frac{3}{8}\beta_{jj}
                           -\frac{1}{2}\delta_{jj}\right) |A_j|^2
\nonumber \\
              +\sum_{l=1(l\ne j)}^{N} g(c_{jl})|A_l|^2\Biggr) A_j\Biggr]
      \exp\left(i\hkj\cdot\vx\right) + c.c. \Biggr\} + \cdots,
\nonumber
\eea
where
\bea
g(c_{jl}) = \frac{3\Gamma_0}{32}\left(1+2c_{jl}^2\right)
          + \frac{7}{8}\left(3-\sqrt{2(1+c_{jl})}-\sqrt{2(1-c_{jl})}\right)
\nonumber \\
          + \left(1+c_{jl}-\sqrt{2(1+c_{jl})}\right)
            \left(\frac{1}{4}w_{jl} - v_{jl}\right)
\nonumber \\
          + \left(1-c_{jl}-\sqrt{2(1-c_{jl})}\right)
            \left(\frac{1}{4}\bar{w}_{jl} - \bar{v}_{jl}\right)
\nonumber \\
          + (1+c_{jl})\left(2\alpha_{jl} 
                        + \frac{3}{8}\beta_{jl} - \frac{1}{2}\delta_{jl}\right)
          + (1-c_{jl})\left(2\bar{\alpha}_{jl} 
                        + \frac{3}{8}\bar{\beta}_{jl} 
                        - \frac{1}{2}\bar{\delta}_{jl}\right).
\eea 
The solvability condition therefore reads,
\be
\label{eq:SWA1}
\frac{\partial A_j}{\partial T} =  \gamma A_j - \left[\gamma g(1)|A_j|^2
              +\gamma\sum_{l=1(l\ne j)}^{N} g(c_{jl})|A_l|^2\right] A_j,
\ee
where $j=1,2, \cdots, N$ and
\be
g(1) = \frac{28+9\Gamma_0}{64} +
2\alpha_{jj}+\frac{3}{8}\beta_{jj}-\frac{1}{2}\delta_{jj}.
\ee
Equation (\ref{eq:SWA1}) is the coupled set of standing wave amplitude 
equations (SWAE) for $A_{j}$, which is the central result of this weakly nonlinear 
analysis.  The generic form of the above set of amplitude equations is 
of course quite general 
and has been derived for a number of different physical systems 
(Cross \& Hohenberg 1993; Newell, Passot \& Lega 1993).
The behavior peculiar to each system stems from the 
functional form of the nonlinear interaction coefficients $g(1)$ and 
$g(c_{jl})$, and from the time constant $\tau_{0}$ ($\tau_0 = 1/\gamma$, in the 
case of Faraday waves.)

Before we look at the quantitative details of the nonlinear coefficients, 
$g(1)$ and $g(c_{jl})$, we have the following comments on the SWAE's.

(i) The exclusion of quadratic nonlinear terms in the standing wave amplitude
equations, which is related to the absence of solvability conditions 
at ${\cal O}(\varepsilon)$, is a consequence of the requirement of sign
invariance of the SWAE's ($A_j \rightarrow -A_j$).  Subharmonic response of
the fluid surface to the driving force $f\sin(2\omega_0 t)$ implies 
$h(\vx, t+\pi/\omega_0) = - h(\vx, t)$, where $h$ is a linear unstable
mode given by Eq.~(\ref{eq:linear_h_x2}).  We note that a sign change of
the amplitude $A_j$ is equivalent to a time displacement in a period of
the driving force, $t \rightarrow t+\pi/\omega_0$.  Because of the
invariance of the original fluid equations under such a time displacement,
the amplitude equation of $A_j$ must be invariant under a sign change in
$A_j$.

(ii) The coefficients of cubic nonlinear terms, $\gamma g(1)$ and $\gamma
g(c_{jl})$, are proportional to the linear damping coefficient $\gamma$.  
This result is
qualitatively different from that obtained by  \cite{re:milner91}.  He
also derived a coupled set of standing wave amplitude equations of the
same form as Eq.~(\ref{eq:SWA1}). Although  his nonlinear coefficients are 
also
proportional to $\gamma$ (the dissipation function $Q$ is after all 
proportional to the kinematic viscosity $\nu$), they
result entirely from nonlinear viscous terms in the dissipation
function. In fact,  linear viscous terms did not contribute at all to 
third order. The appearance of nonlinear terms 
proportional to the linear damping coefficient $\gamma$ in the SWAE's in 
our approach is, however, no surprise.  In general, a parameter that 
appears in the coefficients of the linear terms
in the original equations can appear in the coefficients of
nonlinear terms of the amplitude equations for that system.  For example, 
the nonlinear interaction coefficient in the amplitude equation for
Rayleigh-B\'enard convection is a function of the Prandtl number $Pr$,
although $Pr$ appears only in the coefficients for linear terms in
the Boussinesq equations (see e.g., Cross 1980). Throughout this paper,
nonlinear terms that are proportional to the kinematic viscosity $\nu$ in
the fluid equations (or the quasi-potential equations, Eqs. 
(\ref{eq:laplace_phi})-(\ref{eq:bc_infty})) are termed 
{\em nonlinear viscous terms\/}, and the nonlinear terms proportional to
$\nu$ or $\gamma$ in amplitude equations for Faraday waves are termed 
{\em nonlinear damping terms\/} in order to avoid confusion due to
terminology.  The validity of the quasi-potential equations with only linear
viscous terms relies on the assumption that nonlinear viscous terms do not 
have significant effect on pattern formation in Faraday waves. A check of 
the validity can only be provided by comparing our results to 
experimental studies.

(iii) There are contributions to the nonlinear terms of Eq.~(\ref{eq:SWA1}) 
from the parametric driving force.  These contributions are proportional to
the driving amplitude $f$, but they appear in Eq.~(\ref{eq:SWA1}) together
with the contributions from the linear viscous terms in the quasi-potential
equations since we have set $f=\gamma$ in the linear solutions
(Eqs.~(\ref{eq:order1_h}) and (\ref{eq:order1_Phi})).  The contributions 
from the parametric driving force are directly related to the higher 
harmonic
terms proportional to $f$ in the linear solutions for $h$ and $\Phi$.  This
contribution provides the {\em amplitude-limiting effect} by the driving
force.  This amplitude-limiting effect results from the nonlinear
interaction of these higher harmonic terms with the primary mode, which has 
half of the driving frequency.  Such nonlinear interactions produces terms 
that are out of phase by $\pi/2$ with the primary mode, and thus possibly 
damps or limits the wave amplitude.  An important point 
is that this amplitude-limiting effect results
from non-dissipative terms in the governing equation. Therefore, such an
effect is also important for inviscid systems. This amplitude-limiting 
effect by the driving force, to our knowledge, has not been identified 
before. A yet different amplitude-limiting effect also through non-dissipative
terms was studied by \cite{re:zakharov75} in the context of 
parametric spin-wave systems. Zakharov et al. studied parametric spin-wave
instabilities, and considered the deviation of the phase of the excited spin
waves from the optimum phase as the major nonlinear mechanism which limits 
the parametric instability. However, we agree with 
Cross \& Hohenberg (1993) that this \lq\lq de-phasing" effect of the parametric 
mode is small (of higher order) close to threshold.

Due to the mode interference occurring exactly at $c_{jl} = 1$ ($j=l$ or
$\theta_{jl} = 0$), $g(c_{jl})$ is not a smooth function of $c_{jl}$.  
It is easy to show that
\begin{equation}
\label{eq:nonsmooth_g}
g(1) = \frac{1}{2}g(c_{jl} \rightarrow 1).
\end{equation}
We also note that 
\be
\label{eq:chiral_symmetry}
g(c_{jl}) = g(-c_{jl}).
\ee
This symmetry is an obvious 
requirement for standing wave amplitude equations since it is equivalent to 
have two standing waves separated by angle $\theta$ or by $\pi - \theta$.

An additional issue concerns the transformation of the equations of motion
under time reversal, and the related question of the existence of cubic terms
in Eq.~(\ref{eq:SWA1}) in the limit of a Hamiltonian system (i.e., in the 
absence of dissipative contributions to the equations of motion). Under time
reversal, the variables of interest transform according to $t \rightarrow -t$,
$\hat{k}_{j} \rightarrow - \hat{k}_{j}$ and $A_{j}(t) \rightarrow
A_{j}^{*}(-t)$. We also note that the Hamiltonian 
Eq.~(\ref{eq:hamiltonian_explicit}) depends explicitly on time, and that 
given our choice of driving force proportional to $\sin 2t$,
the trajectory of the system under time reversal is
invariant only if, in addition to the transformation rules given above, 
$f \rightarrow -f$. A similar situation arises in systems with applied 
magnetic fields, or for Coriolis forces, in which either the magnetic field
or the angular velocity must be reversed. In the absence of viscous dissipation
($\gamma=0$), the only nonlinear terms in the SWAE's come from 
the driving force, and are proportional to the driving amplitude $f$.  
These terms can be written as, 
$$
-f\left[g(1)|A_j|^2
+\sum_{l=1(l\ne j)}^{N} g(c_{jl})|A_l|^2\right] A_j.
$$
Thus, nonlinear terms proportional to $f$ do change sign under time reversal,
and are allowed in the equations of motion for the amplitudes $A_{j}$. This
conclusion is not trivially related to having chosen a sine
function as the driving force.  Had we chosen $f\cos(2t)$ as the
forcing term, the linearly growing modes would be a combination of $A_j$ and 
$B_j$, the algebra for the derivation of the SWAE's becomes more tedious, but 
the final conclusion remains the same even though, in this latter case, $f$ 
would be formally chosen to be invariant under time reversal.
On the other hand, for an autonomous system in which the Hamiltonian
does not explicitly depends on time, if the cubic term in Eq.~(\ref{eq:SWA1}) 
is entirely of Hamiltonian character, then invariance under time reversal 
implies $\sum_{l} g(c_{jl}) |A_{l}|^{2}A_{j}=0$, for any arbitrary set of 
amplitudes. This equality is satisfied if $g(c_{jl}) = - g(-c_{jl})$, 
where $c_{jl}$ involves the interaction between modes $\hat{k}_{j}$ and 
$\hat{k}_{l}$, and $-c_{jl}$ between $\hat{k}_{j}$ and $-\hat{k}_{l}$ 
(see Fig.~\ref{fig-triad} and Cross \& Hohenberg (1993)). This symmetry
together with Eq.~(\ref{eq:chiral_symmetry}) would imply that 
$g(c_{jl})=0$ in this case. Therefore, saturation of the wave amplitude
would have to occur either through weak nonlinear dissipative effects, or 
higher order terms (e.g., nonlinear frequency detuning terms).

\begin{figure}
\vspace{2.5in}
\includegraphics{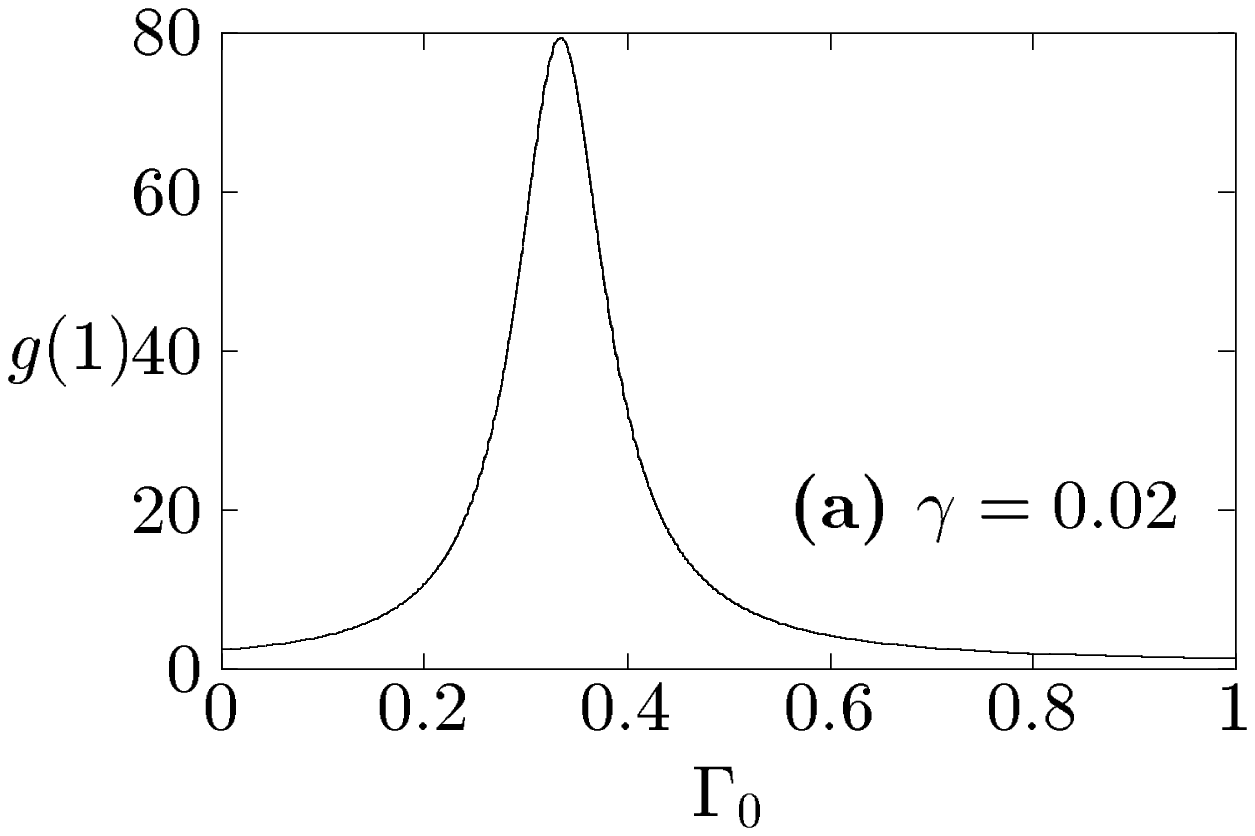}
\includegraphics{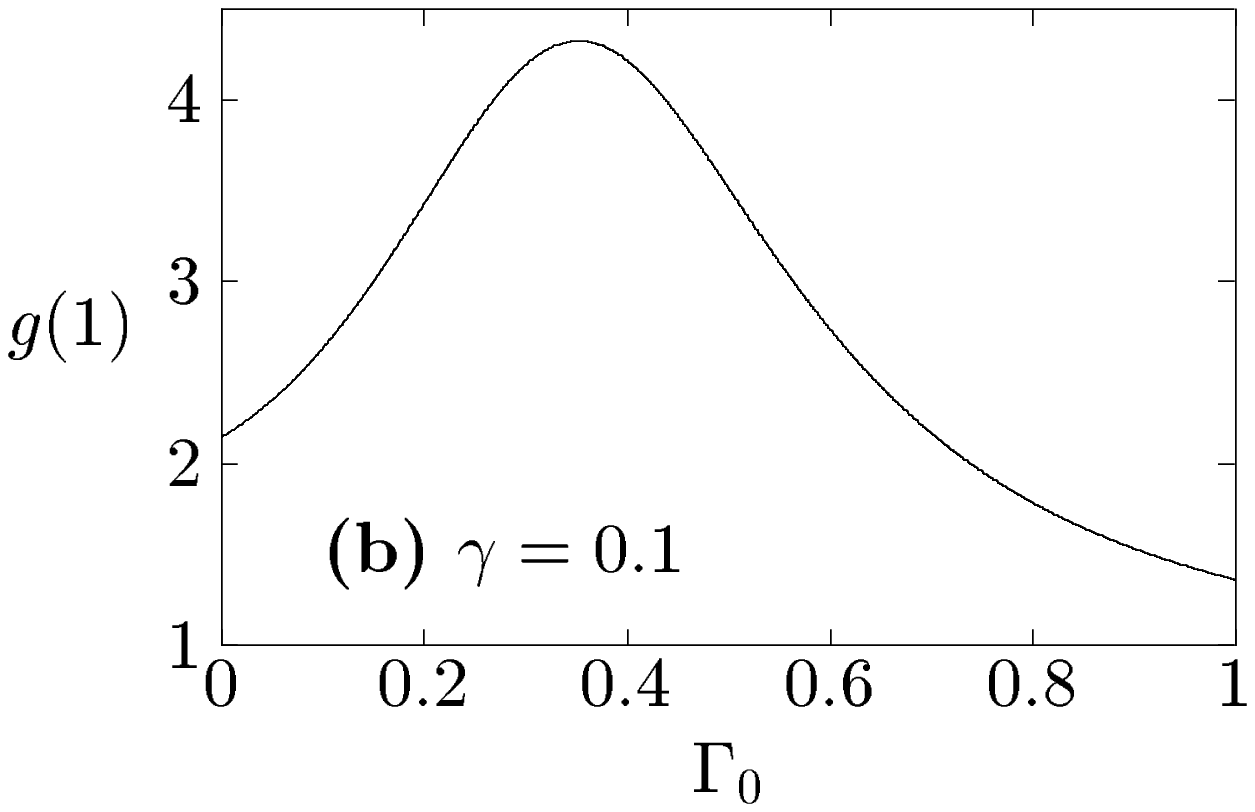}
\caption{The self-interaction coefficient $g(1)$ of the standing wave amplitude
equations as a function of $\Gamma_0$ with the linear damping coefficient 
$\gamma=0.1$ in (a), and $\gamma=0.02$ in (b).}
\label{fig-g1Gamma}
\end{figure}

Figure \ref{fig-g1Gamma} 
shows the self-interaction coefficient $g(1)$ as a function of $\Gamma_0$ 
for two different values of $\gamma = 0.1$ and $0.02$.  The maxima in the 
curves around $\Gamma_0 = 1/3$ are the result of the triad resonant 
interaction of standing waves in the same direction: $\theta_{jl}^{(r)} = 0$
(the second-harmonic resonance), that affect $g(1)$.  Since the resonance is
less damped for smaller values of $\gamma$, the value of the coefficient 
$g(1)$ varies inversely with damping.

Since $g(1) > 0$, we 
can rescale the amplitude as $\tilde{A}_j = \sqrt{g(1)}A_j$.  We have the 
following standing wave amplitude equation for the scaled amplitude,
\be
\label{eq:SWA2}
\frac{1}{\gamma}\frac{\partial \tilde{A}_j}{\partial T} =  \tilde{A}_j
- \left[|\tilde{A}_j|^2 +\sum_{l=1(l\ne j)}^{N} \tilde{g}(c_{jl})
|\tilde{A}_l|^2 \right] \tilde{A}_j,
\ee 
where $\tilde{g}(c_{jl})=g(c_{jl})/g(1)$. From Eq.~(\ref{eq:nonsmooth_g})
and symmetry for $\pm c_{jl}$, we also have 
$\tilde{g}(c_{jl} \rightarrow \pm 1) = 2$.  

\begin{figure}
\vspace{5.0in}
\includegraphics{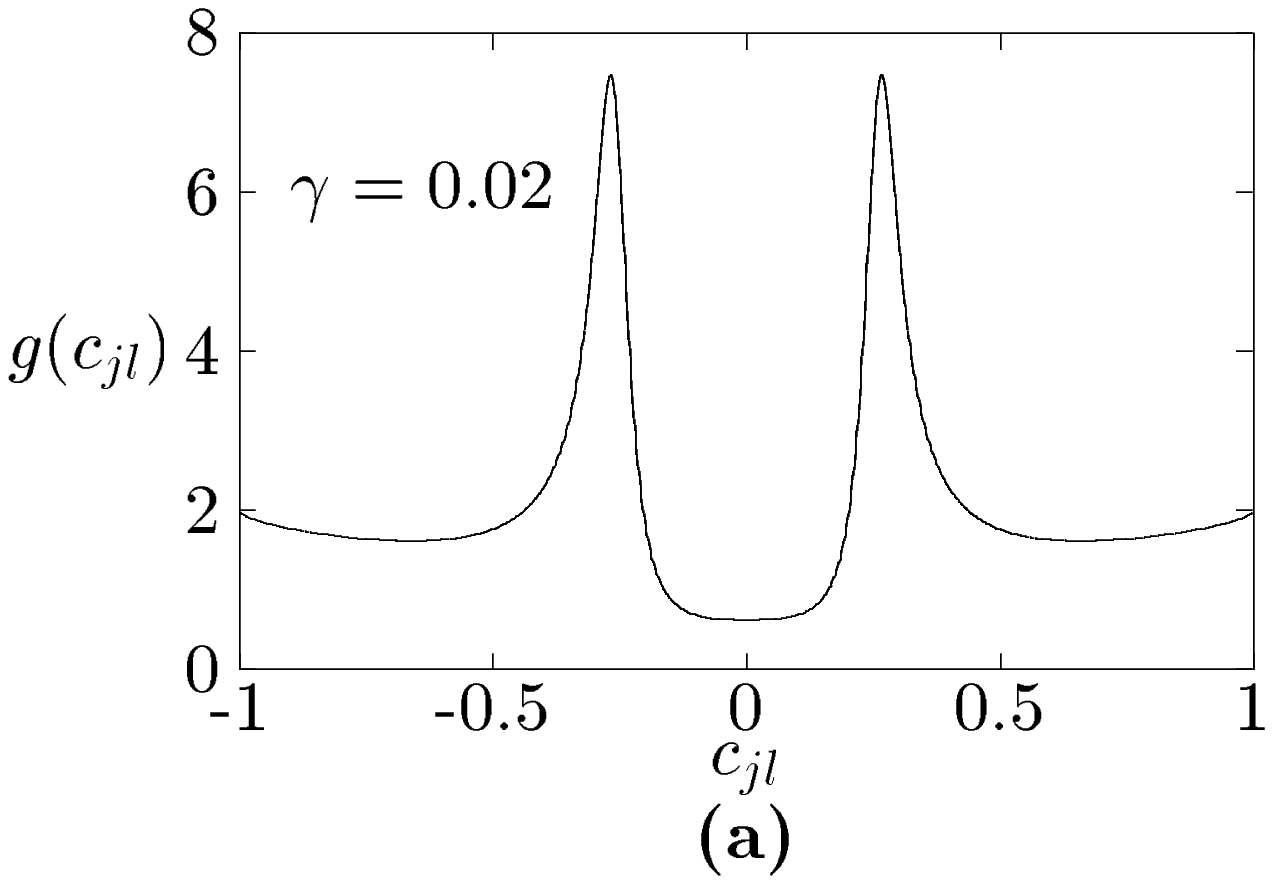}
\includegraphics{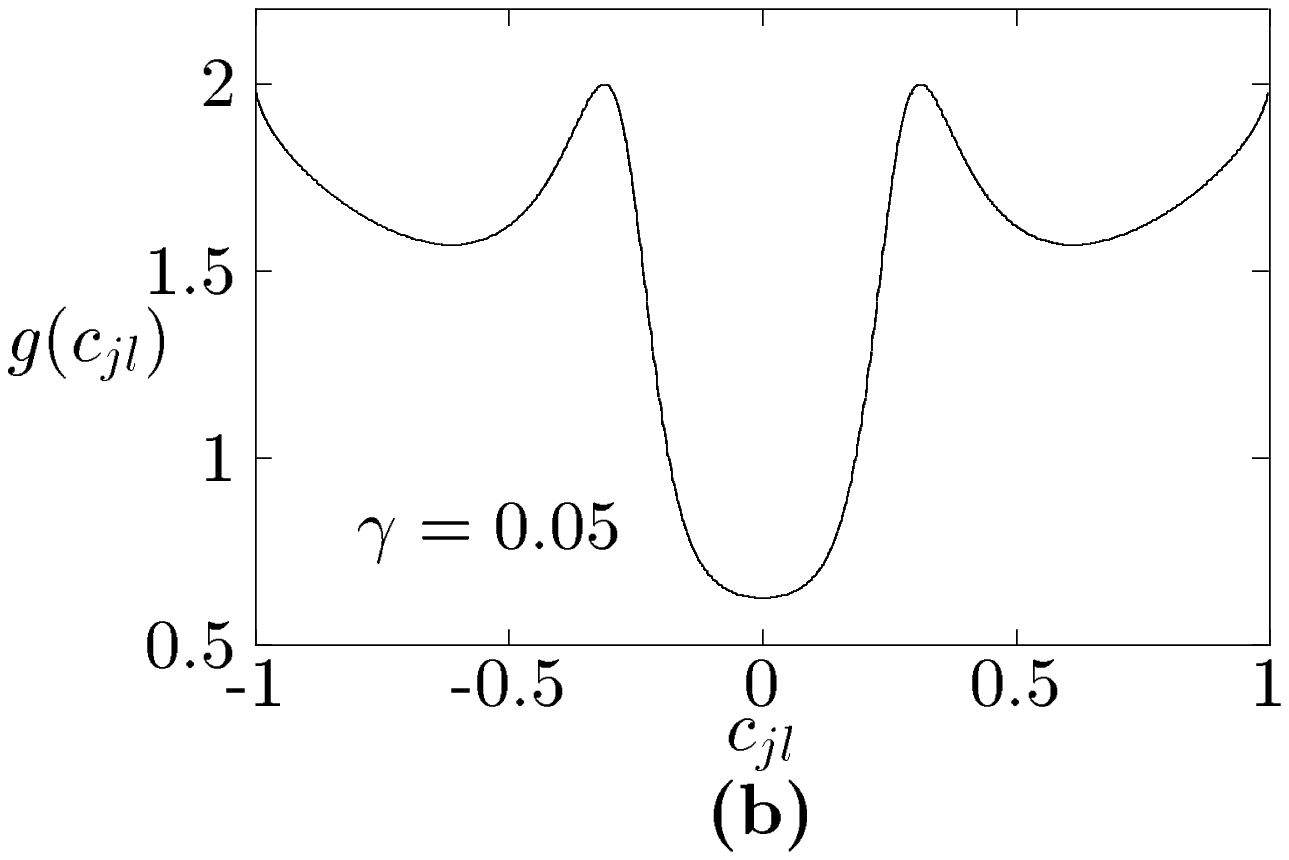}
\includegraphics{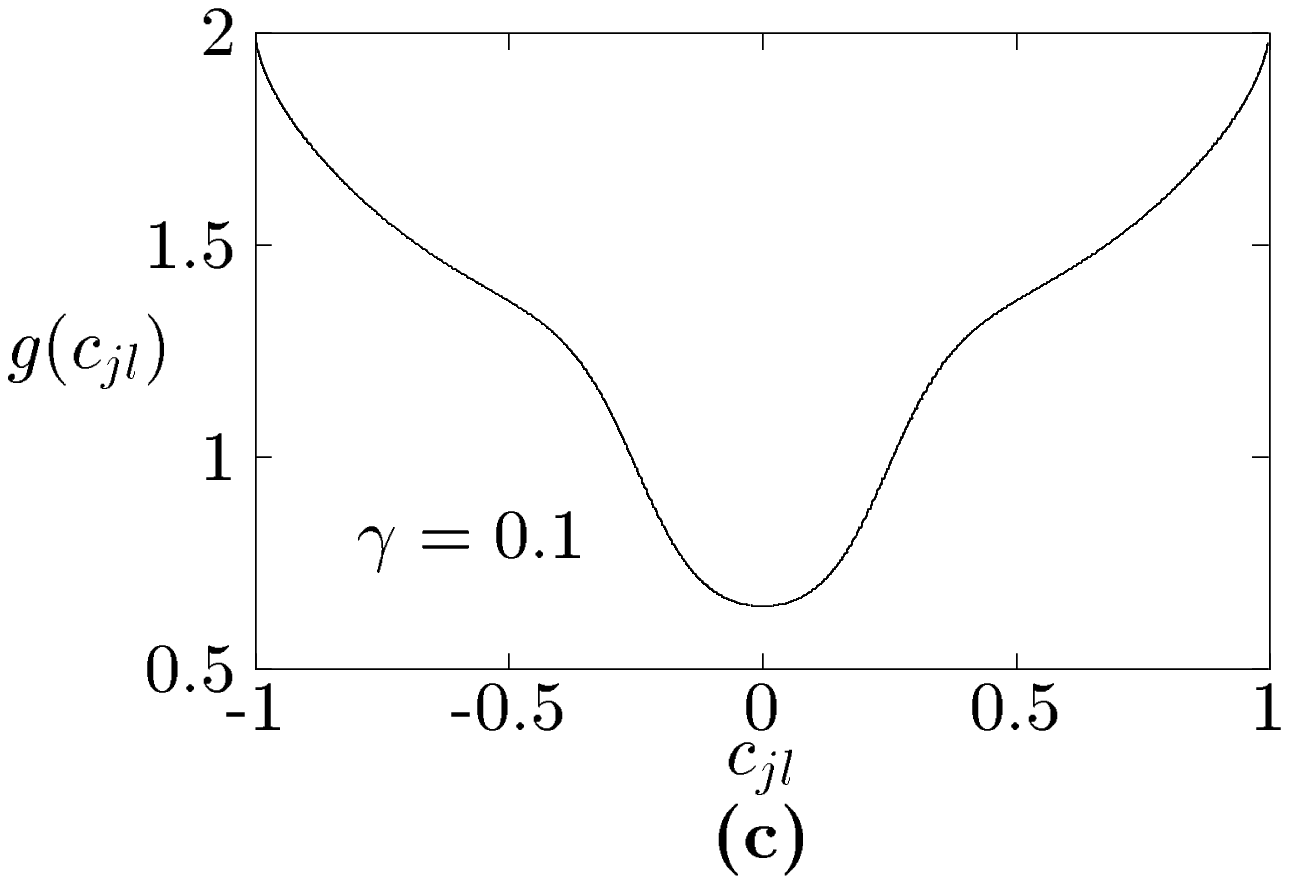}
\includegraphics{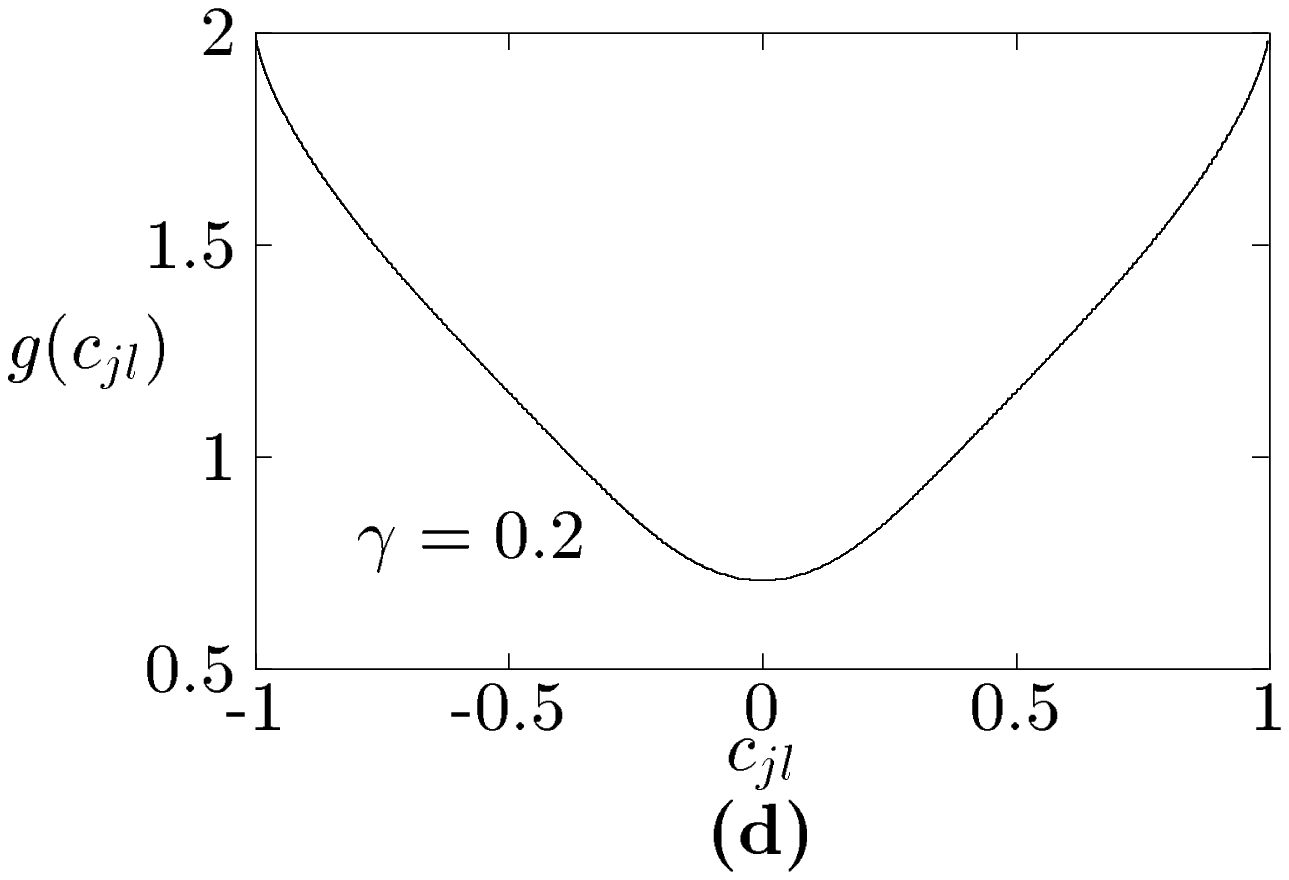}
\caption{The standing wave nonlinear coefficient $g(c_{jl})$ as a
function of $c_{jl}$ for purely capillary waves ($\Gamma_0=1$) with the 
linear damping coefficient $\gamma=0.02$ in (a), $\gamma=0.05$ in (b), 
$\gamma=0.1$ in (c), and $\gamma=0.2$ in (d).}
\label{fig-gTheta1.0}
\end{figure}

The nonlinear interaction coefficient $g(c_{jl})$ in Eq.~(\ref{eq:SWA2}) 
(we have suppressed the tilde since we will only refer to the scaled nonlinear 
coefficient in what follows) depends on the dimensionless fluid parameters 
$\Gamma_0$ (or $G_0$) and $\gamma$.  Figure \ref{fig-gTheta1.0}
shows the function $g(c_{jl})$ for four different values
of the damping coefficient $\gamma$.  The maxima in $g(c_{jl})$ around 
$c_{jl}=0.26$ ($\theta_{jl} = 74.9^{\circ}$) correspond to triad 
resonance for purely capillary waves (see Fig.~\ref{fig-triad}).
Even for relatively large values of the damping coefficient $\gamma=0.2$, the 
influence of the triad resonance on the $g(c_{jl})$ curve can still be seen,
but becomes much weaker.  An important feature common to all the curves in
Fig.~\ref{fig-gTheta1.0} is that there is a minimum of $g(c_{jl})$ at 
$c_{jl}=0$,
and $g(0) < 1$.  It is also 
interesting to compare the differences in $g(c_{jl})$ for 
purely capillary waves ($\Gamma_0 = 1$), purely gravity waves ($\Gamma_0 = 
0$), 
and mixed gravity-capillary waves ($0 < \Gamma_0 < 1$) since the triad 
resonant 
interaction strongly depends on the value of $\Gamma_0$ .  
Figure \ref{fig-gTheta0.0}
shows the function $g(c_{jl})$ for four different values of damping 
coefficients 
(the same as in Fig.~\ref{fig-gTheta1.0}) for purely gravity waves 
($\Gamma_0 = 0$).  
We note that because of the absence of triad resonant interaction,
the variations among the $g(c_{jl})$ curves for different values of $\gamma$ 
are quite small.  The curves still have minima at $c_{jl} =0$, but 
the minima are much flatter.

Since the triad resonant interaction occurs among waves with their wavevectors
in the same direction when $\Gamma_0 = 1/3$, we finally examine the 
function $g(c_{jl})$ for this case.  Figure \ref{fig-gTheta13}
shows $g(c_{jl})$ for four different values of the damping coefficient (the 
same
as in Fig.~(\ref{fig-gTheta1.0})) for capillary-gravity waves with 
$\Gamma_0 = 1/3$.  
We observe that $g(c_{jl})$ is very flat and reaches very small positive 
values for small values of $\gamma$.  These facts are a consequence of the 
second-harmonic resonance, since the value of $g(1)$ becomes very large (see
Figure \ref{fig-g1Gamma}) for this special case of triad resonant interaction.
For relatively large values of the damping parameter, e.g., $\gamma=0.20$, the 
effect of the second-harmonic resonance is much smaller.

\subsection{Pattern Selection Near Onset}
\label{sec:SWAE-selection}

Equation (\ref{eq:SWA2}) is of gradient form 
$1/\gamma \partial_T A_j = -\partial \sF/\partial A_j^*$, with Lyapunov 
function 
$\sF$ given by,
\be
\sF = - \sum_{j=1}^{N} |A_j|^2 + \frac{1}{2}\sum_{j=1}^{N} |A_j|^2
     \left(|A_j|^2 + \sum_{l=1(l\ne j)}^N g(c_{jl}) |A_l|^2\right).
\ee
Since
\be
\frac{d\sF}{dT} = \sum_{j=1}^{N} \left(\frac{\partial \sF}{\partial A_j}
                                       \partial_T A_j
                                      +\frac{\partial \sF}{\partial A_j^*}
                                       \partial_T A_j^* \right)
                = - \frac{2}{\gamma}\sum_{j=1}^{N} |\partial_T A_j|^2 \le 0,
\ee
the only possible limiting cases of such a dissipative system, in the limit 
$T \rightarrow \infty$, are stationary states for the amplitudes $A_j$.  Only 
the states which correspond to local minima of the Lyapunov function are 
linearly stable.

Apart from the trivial solution of $A_j =0$ for $j=1, \cdots, N$,
Eq.~(\ref{eq:SWA2}) has a family of stationary solutions differing in the 
total number of standing waves $N$ for which $A_j \ne 0$.  By considering
the case in which the magnitudes of all standing waves are the same, i.e.
$|A_j| = |A|$, Eq.~(\ref{eq:SWA2}) has the following solutions,
\be
\label{eq:SWA_solution}
|A_j| = |A| = \left(1+\sum_{l=1(l\ne j)}^{N} g(c_{jl})\right)^{-1/2}.
\ee
The values of Lyapunov function for these solutions are,
\be
\sF = - \frac{N}{2} |A|^2 = - \frac{N/2}{1+\sum_{l=1(l\ne j)}^{N} g(c_{jl})}.
\ee
We note that the greater the square of the amplitude, the lower value of the 
Lyapunov function.  Also the larger the values of $g(c_{jl})$ for a standing
wave pattern, the larger value of the Lyapunov function.  In particular, if
an angle separating the wavevectors of two standing wave modes of the pattern 
satisfies the triad resonant condition, the corresponding $g(c_{jl})$, and thus
the value of the Lyapunov function, will be large.  Therefore, such patterns
will not likely to appear.  This result is consistent with our intuitive
understanding of the role of the triad resonant interaction, i.e., the system
tries to avoid pairs of standing waves with their wavevectors separated by
an angle satisfying the triad resonant condition.

\begin{figure}
\vspace{2.6in}
\includegraphics{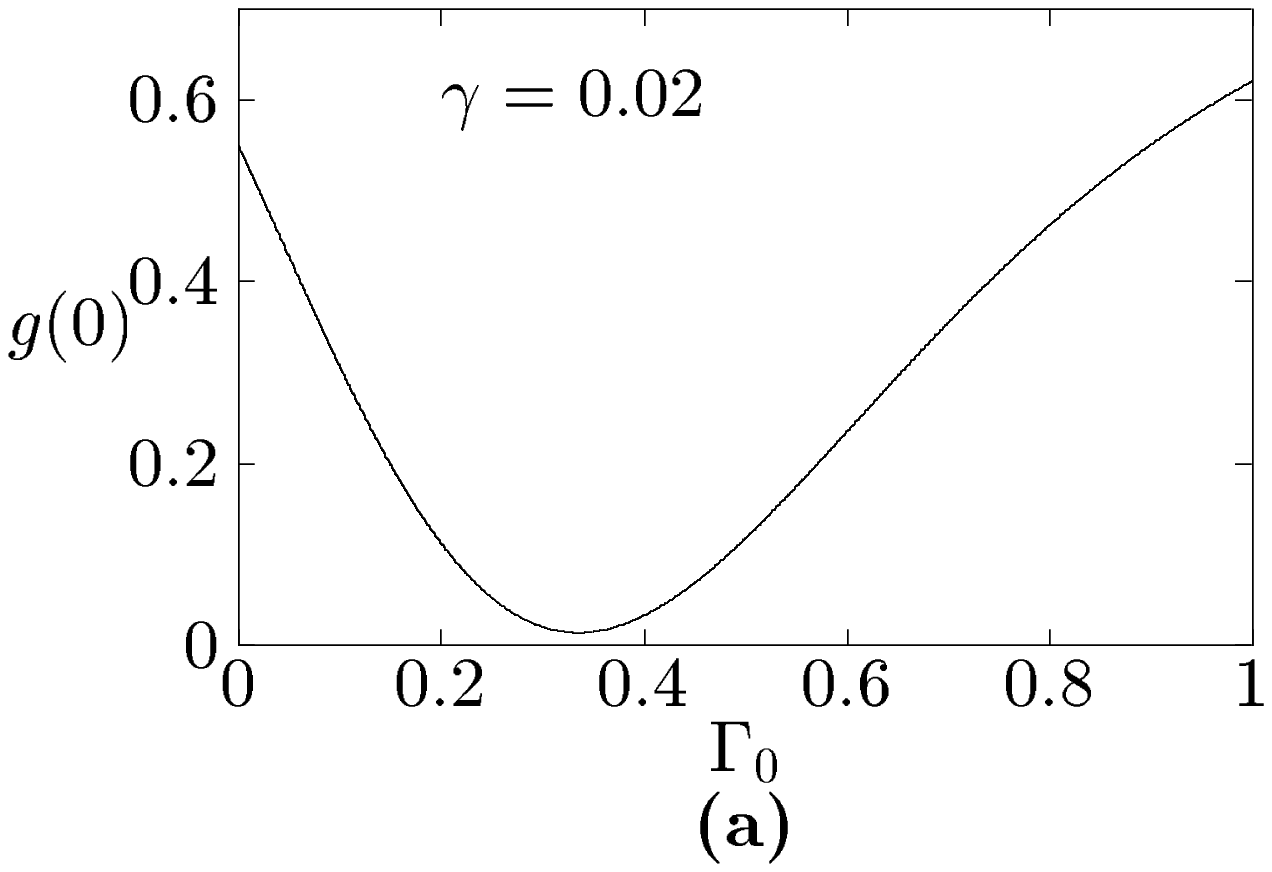}
\includegraphics{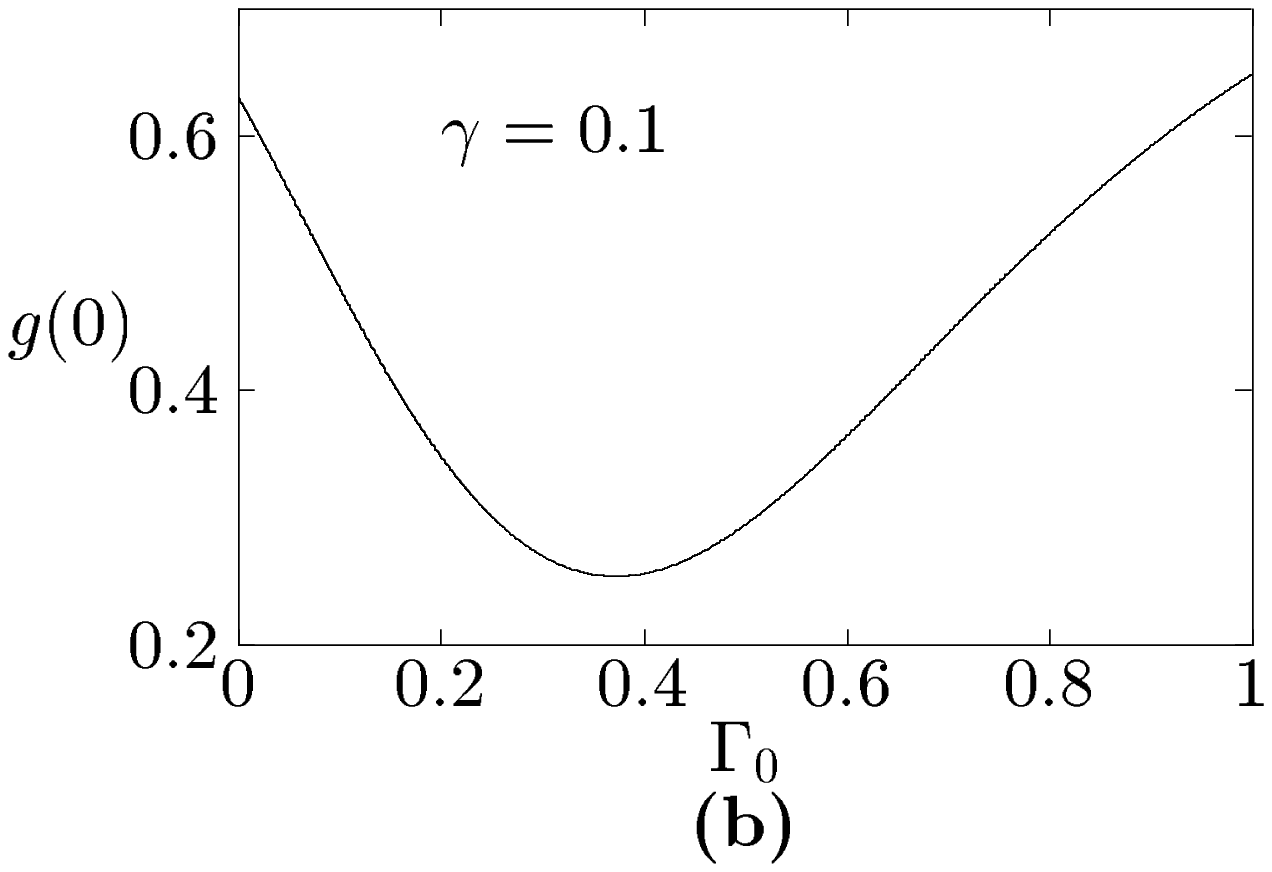}
\caption{$g(0)$ as a function of $\Gamma_0$ for $\gamma = 0.1$ in (a) and 
$\gamma = 0.02$ in (b). The minima around $\Gamma_0 = 1/3$ are the consequence
of the one-dimensional triad resonant interaction.}
\label{fig-g0Gamma}
\end{figure}

For $N=1$ (parallel roll solution), $\sF_1 = -\frac{1}{2}$.  For $N=2$, we have
either square ($c_{12} = 0$) or rhombic ($c_{12} \ne 0$) patterns with 
$\sF_2 = -1/(1+g(c_{12}))$.  By considering regular patterns\footnote{By 
regular patterns, we mean pattern structures for which the angle between 
any two adjacent wavevectors $\vk_j$ and $\vk_{j+1}$ is the same and amounts 
to $\pi/N$.} only, for $N=3$,
we have either hexagonal or triangular patterns, which have the same value of
the Lyapunov function 
$$ 
\sF_3 = - \frac{3/2}{1+g(1/2)+g(-1/2)}.
$$ 
We first consider square patterns for $N=2$.  If $g(0) < 1$, we have 
$\sF_2 < \sF_1 = -\frac{1}{2}$.  As shown in Figure \ref{fig-g0Gamma}
we indeed have $g(0) < 1$ for the interesting parameter range of $\Gamma_0$ 
and $\gamma$.  Therefore standing wave patterns of square symmetry always have
{\em lower} values of Lyapunov function than parallel roll patterns for
weakly damped parametric surface waves near onset.

\begin{figure}
\vspace{2.6in}
\includegraphics{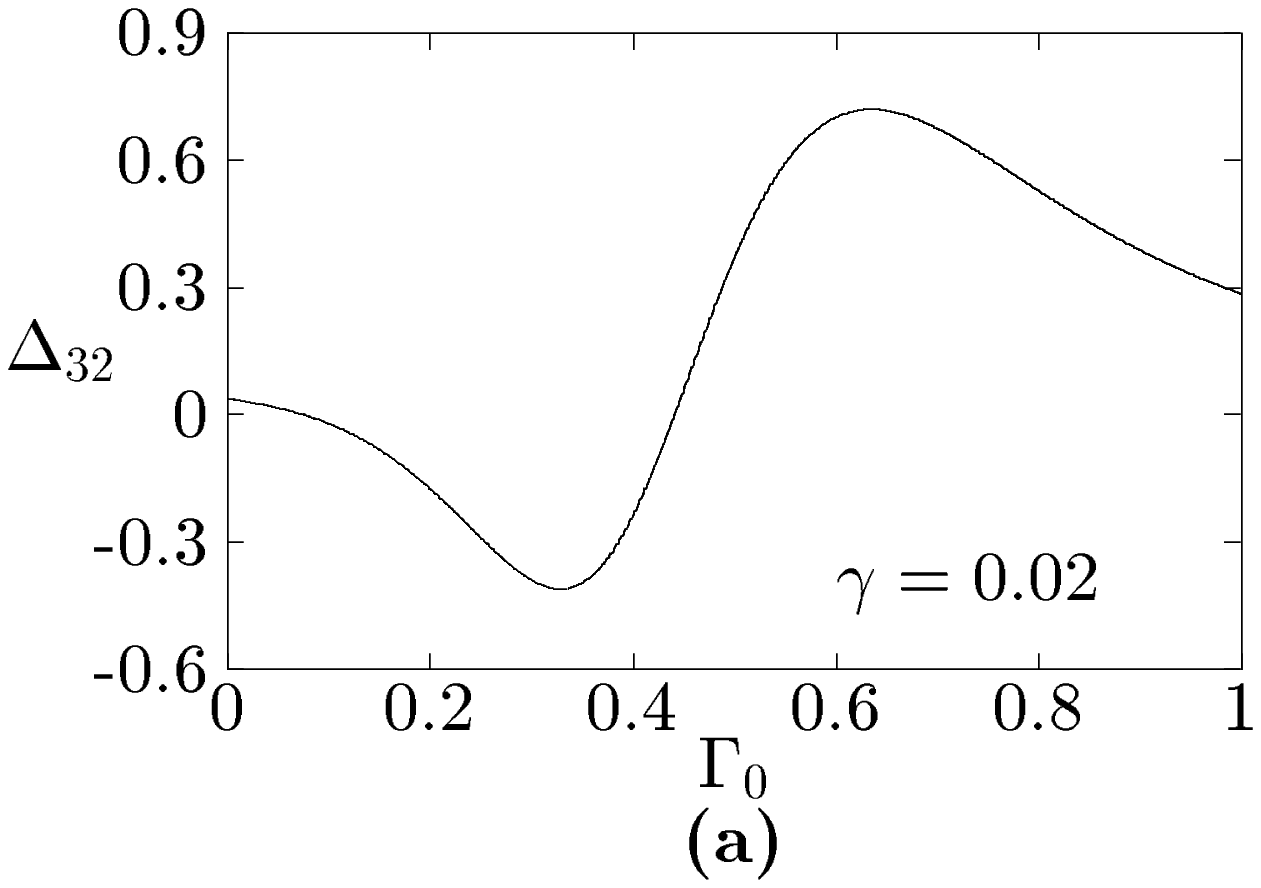}
\includegraphics{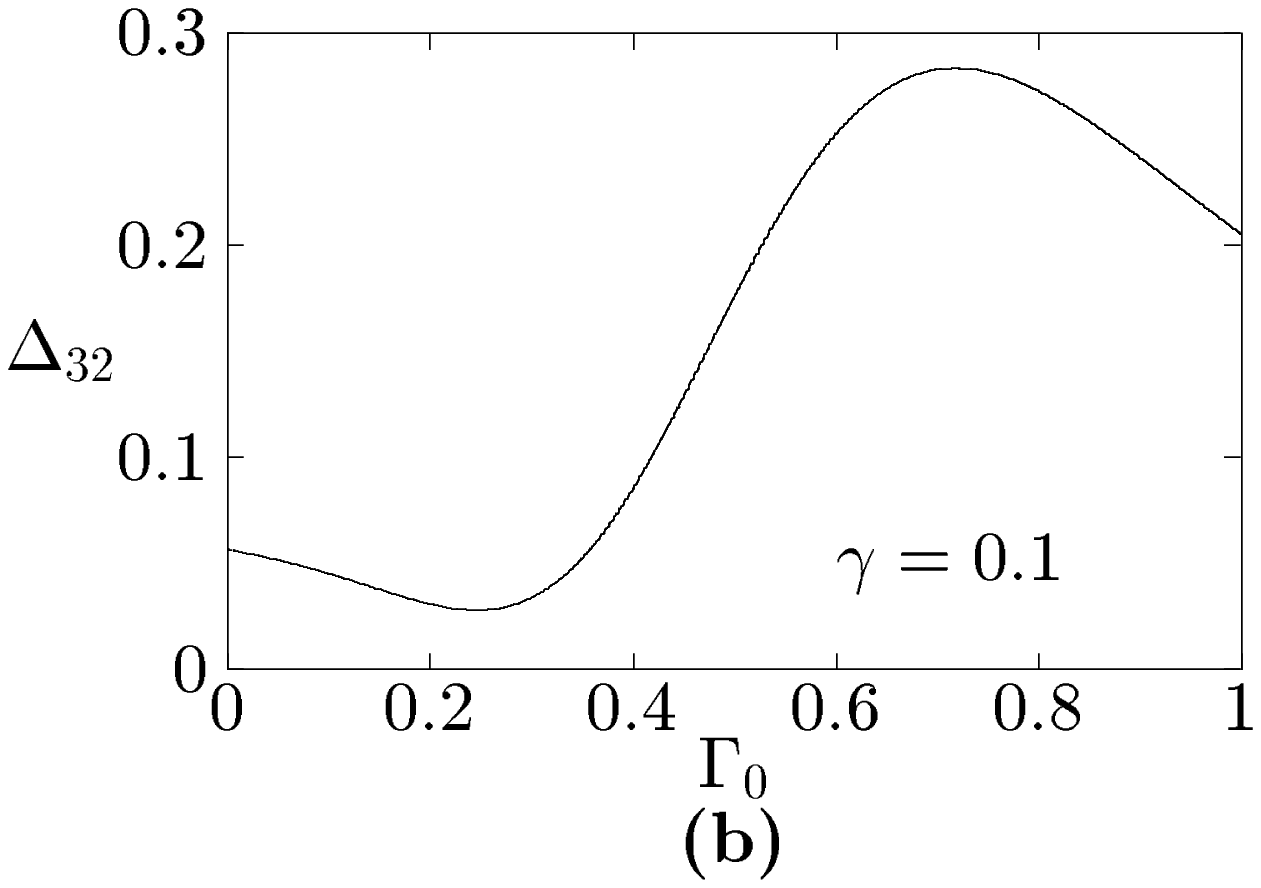}
\caption{The difference $\Delta_{32}$ of the values of the Lyapunov function 
for
hexagonal/triangular patterns $\sF_3$ and square patterns $\sF_2$ as a function
of $\Gamma_0$ for $\gamma = 0.1$ in (a) and $\gamma = 0.02$ in (b).}
\label{fig-Lyapunov3-2}
\end{figure}

In order to compare the values of Lyapunov function for square patterns with
that of hexagonal or triangular patterns, we compute the value of 
$$
\Delta_{32} \equiv \sF_3 - \sF_2 = \frac{1+g(1/2)+g(-1/2) - 
\frac{3}{2}(1+g(0))}
                     {(1+g(0))(1+g(1/2)+g(-1/2))},
$$
which is plotted in Figure \ref{fig-Lyapunov3-2}.

For $\gamma = 0.1$, we have $\Delta_{32} = \sF_3 - \sF_2 > 0$ for all values of
$\Gamma_0$, and thus standing wave patterns of square symmetry also have 
{\em lower} values of the Lyapunov function than hexagonal/triangular 
patterns.  
We also note that the difference between $\sF_3$ and $\sF_2$ becomes smaller 
for smaller values of $\Gamma_0$.  For $\gamma = 0.02$, we still have 
$\Delta_{32} = \sF_3 - \sF_2 > 0$ for capillary waves, but near the 
second-harmonic resonance ($\Gamma_0 = 1/3$),
we have $\sF_3 < \sF_2$, i.e., hexagonal/triangular patterns have lower values 
of the Lyapunov function than square patterns.

\begin{figure}
\vspace{2.6in}
\includegraphics{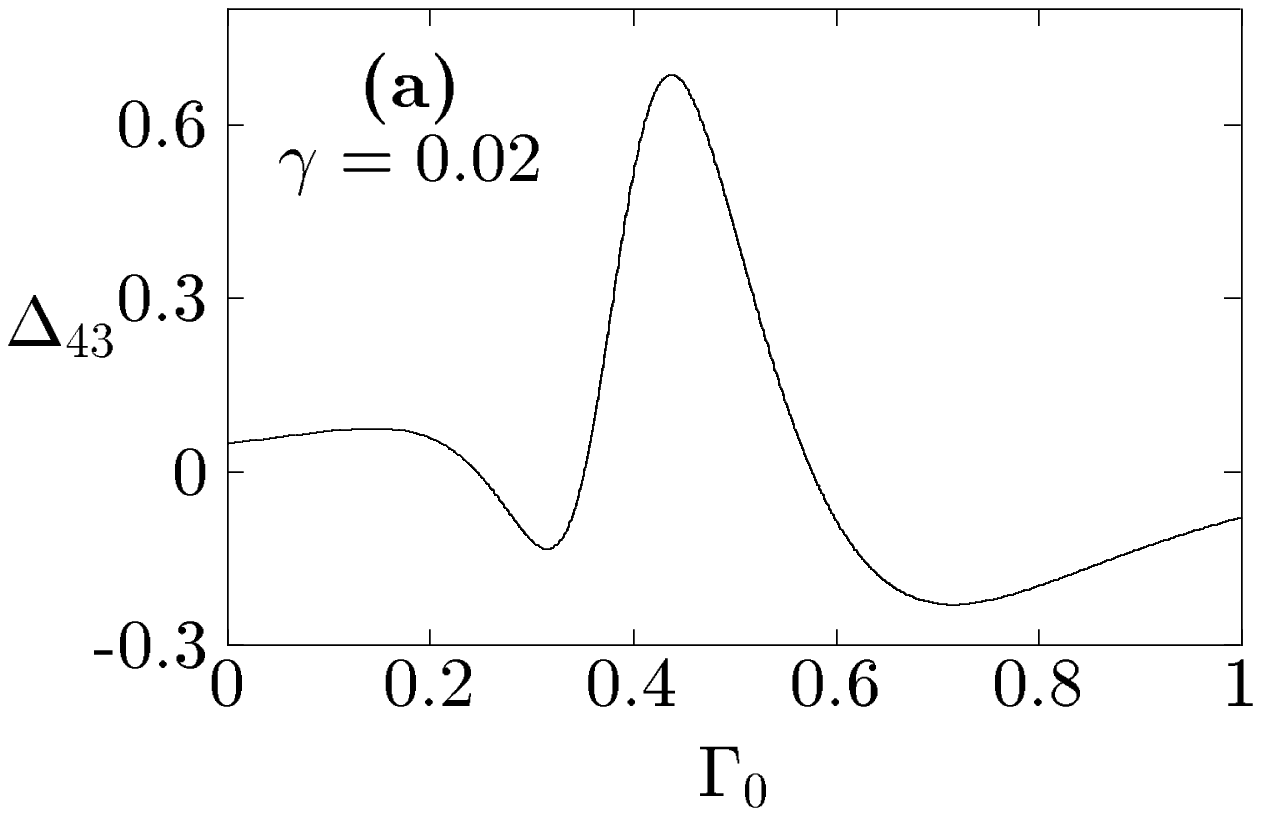}
\includegraphics{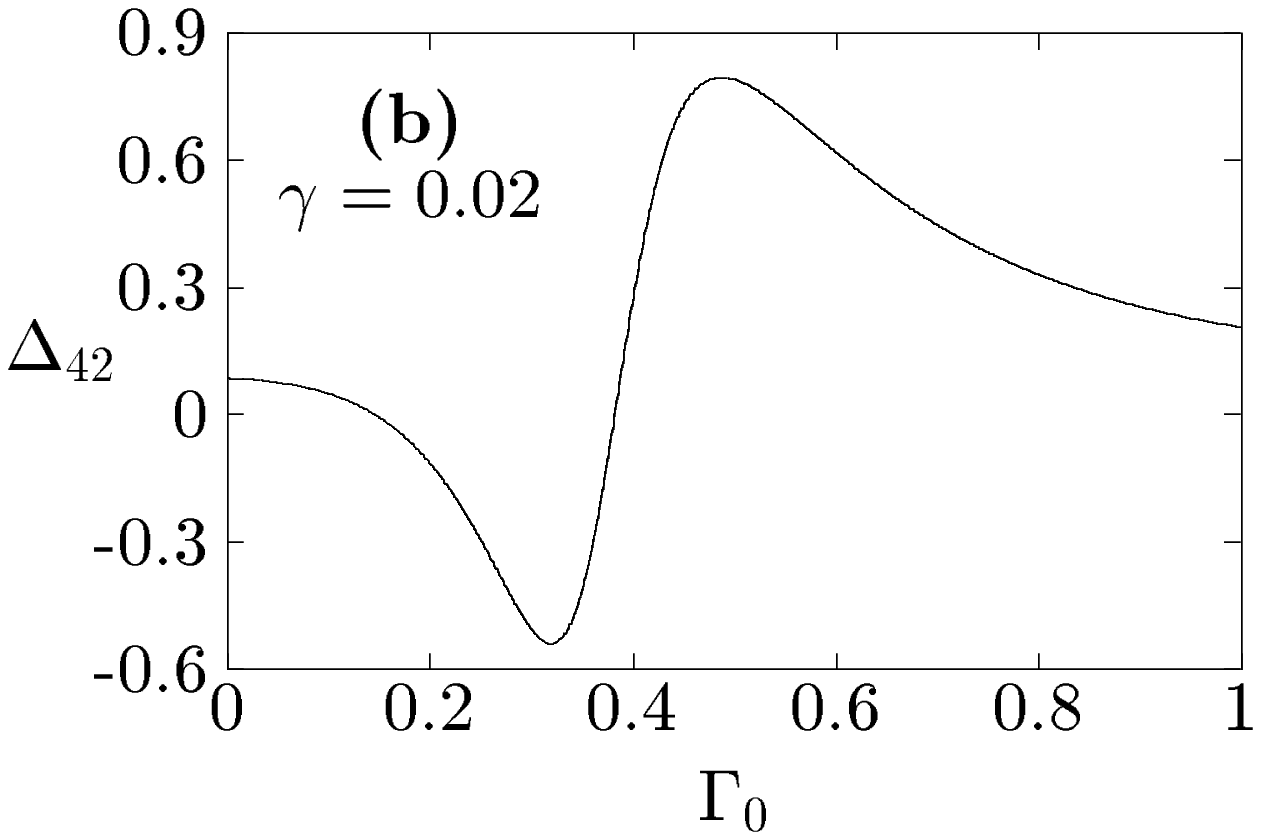}
\caption{The differences in the values of the Lyapunov function for eightfold
quasipatterns and hexagonal/triangular patterns, $\Delta_{43}$, in (a), and
for eightfold quasipatterns and square patterns, $\Delta_{42}$, in (b).  In
both cases, $\gamma=0.02$.}
\label{fig-Lyapunov4-3_4-2}
\end{figure}

Regular patterns for $N \ge 4$ are two-dimensional quasicrystalline patterns 
(or quasipatterns (Edwards \& Fauve 1993, 1994)).
A quasipattern has long-range
orientational order but no spatial periodicity, thus analogous to a 
quasicrystal
in solid state physics (\cite{re:shechtman84}).  For $N = 4$, the value of the 
Lyapunov function for an eightfold quasipattern is,
$$ 
\sF_4 = -\frac{2}{1+g(\sqrt{2}/2) + g(0) + g(-\sqrt{2}/2)}.
$$ 
We are interested in any parameter range in which $\sF_4$ has lower value
of the Lyapunov function than $\sF_2$ and $\sF_3$.  The most possible 
parameter range is certainly near the second-harmonic triad resonant 
interaction with very weak damping.  We thus compute the values of
$$ 
\Delta_{43} \equiv \sF_4 - \sF_3 =
\frac{\frac{3}{2}(1+g(0)) + 3g(\sqrt{2}/2) - 2(1+2g(1/2))}
     {((1+2g(1/2))(1+g(0)+2g(\sqrt{2}/2))},
$$ 
and
$$ 
\Delta_{42} \equiv \sF_4 - \sF_2 =
\frac{2g(\sqrt{2}/2) - (1+g(0))}
     {(1+g(0))(1+g(0)+2g(\sqrt{2}/2))},
$$ 
which are plotted in Fig.~\ref{fig-Lyapunov4-3_4-2}.
We see indeed that for $\gamma=0.02$ we have $\sF_4 < \sF_3$ and $\sF_4 <
\sF_2$ around $\Gamma_0 = 1/3$.

\begin{figure}
\vspace{5.0in}
\includegraphics{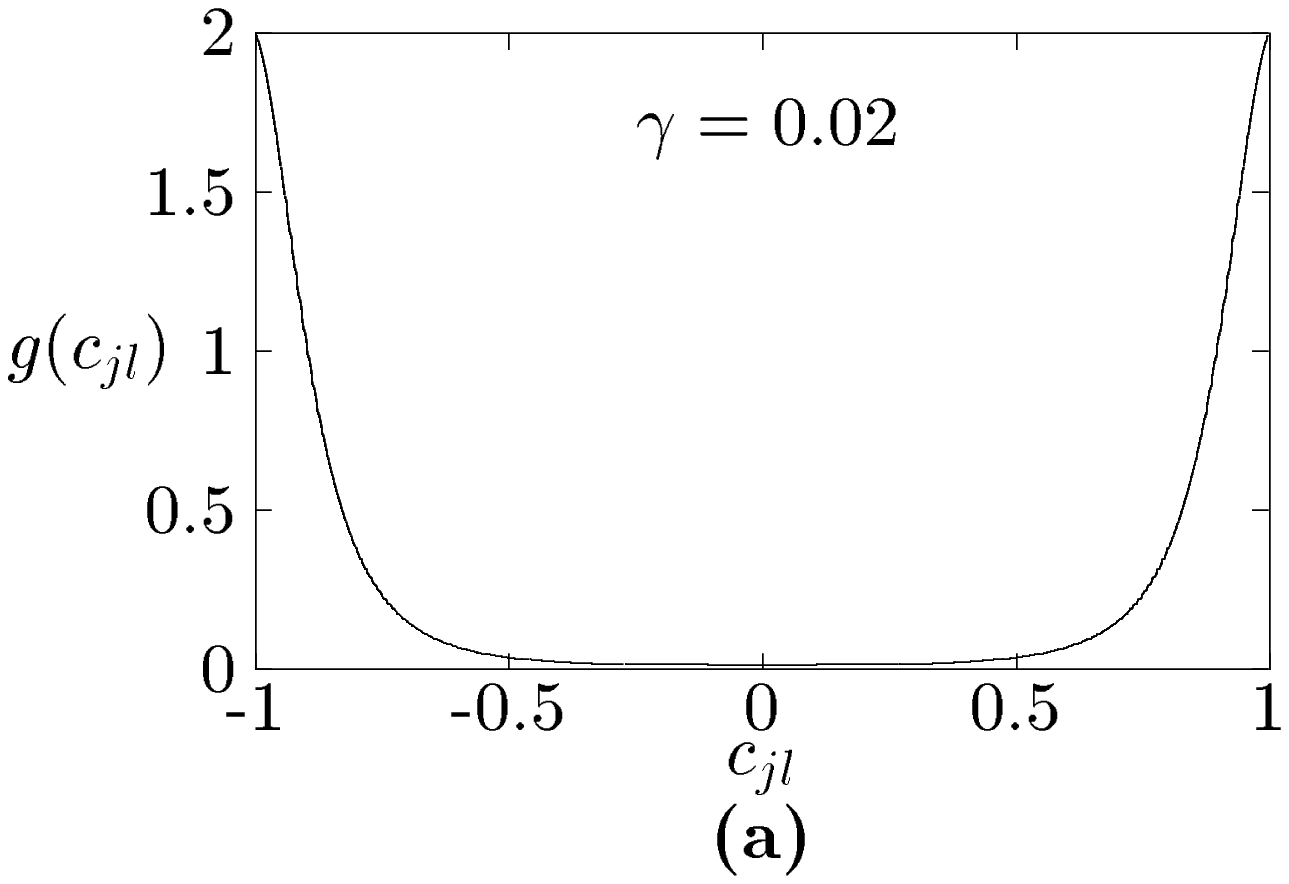}
\includegraphics{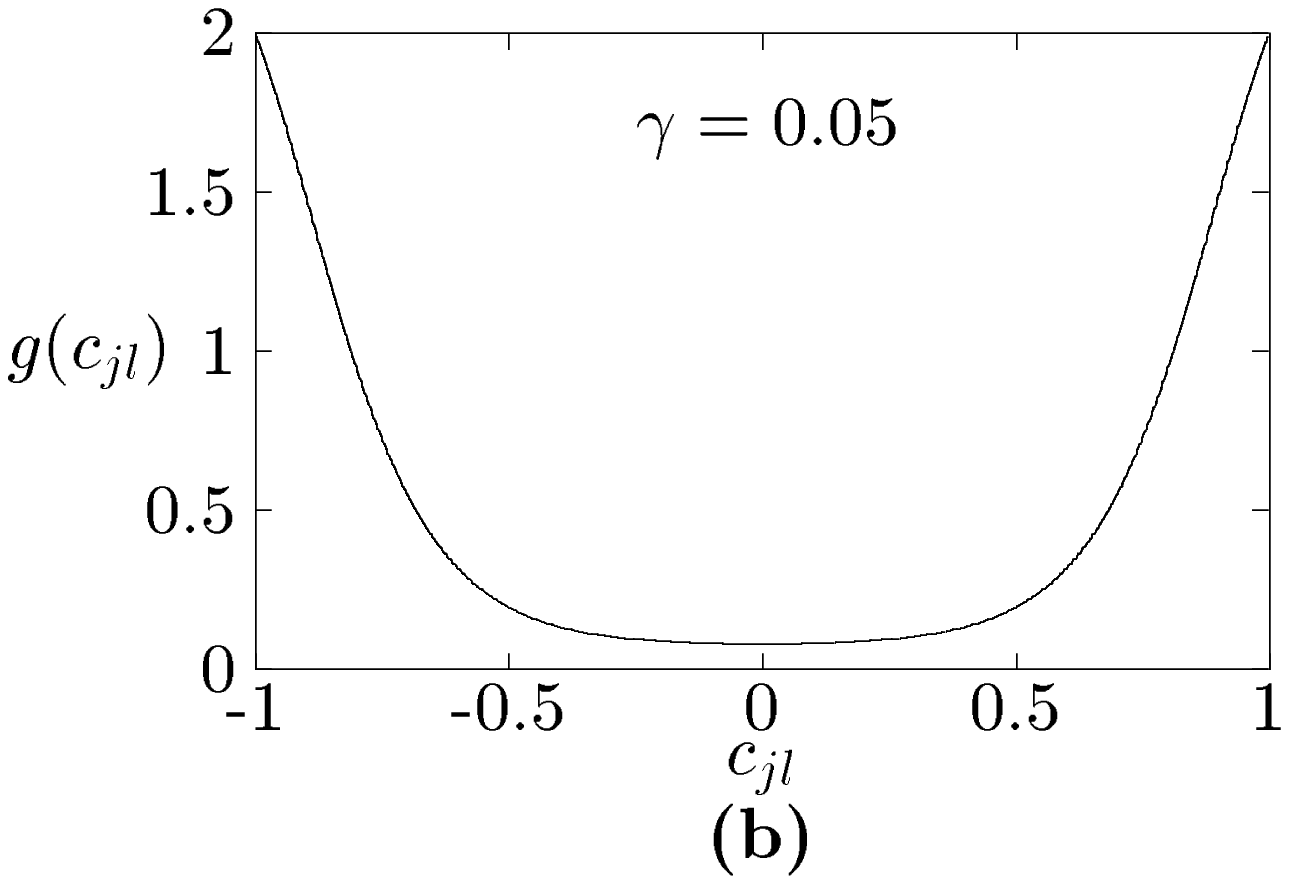}
\includegraphics{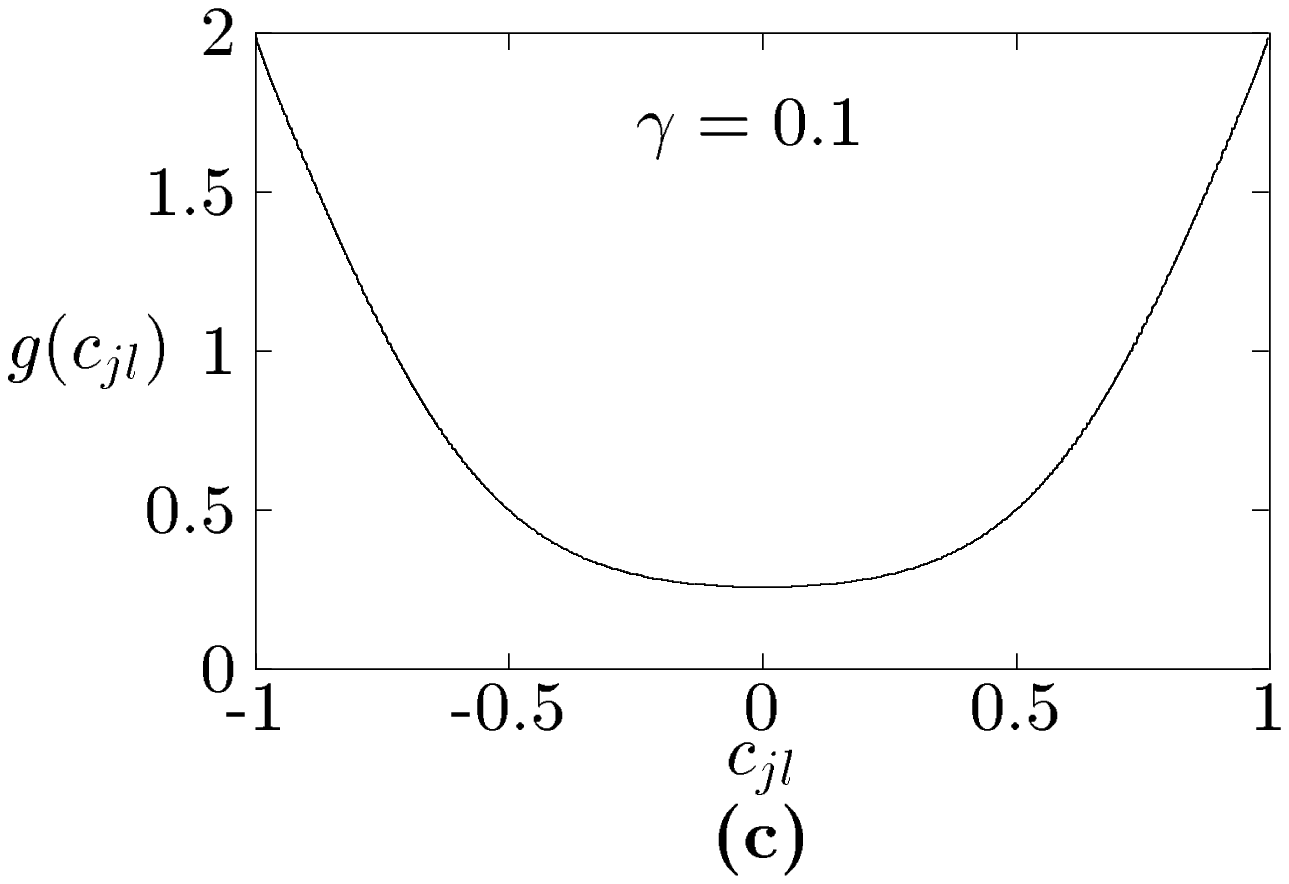}
\includegraphics{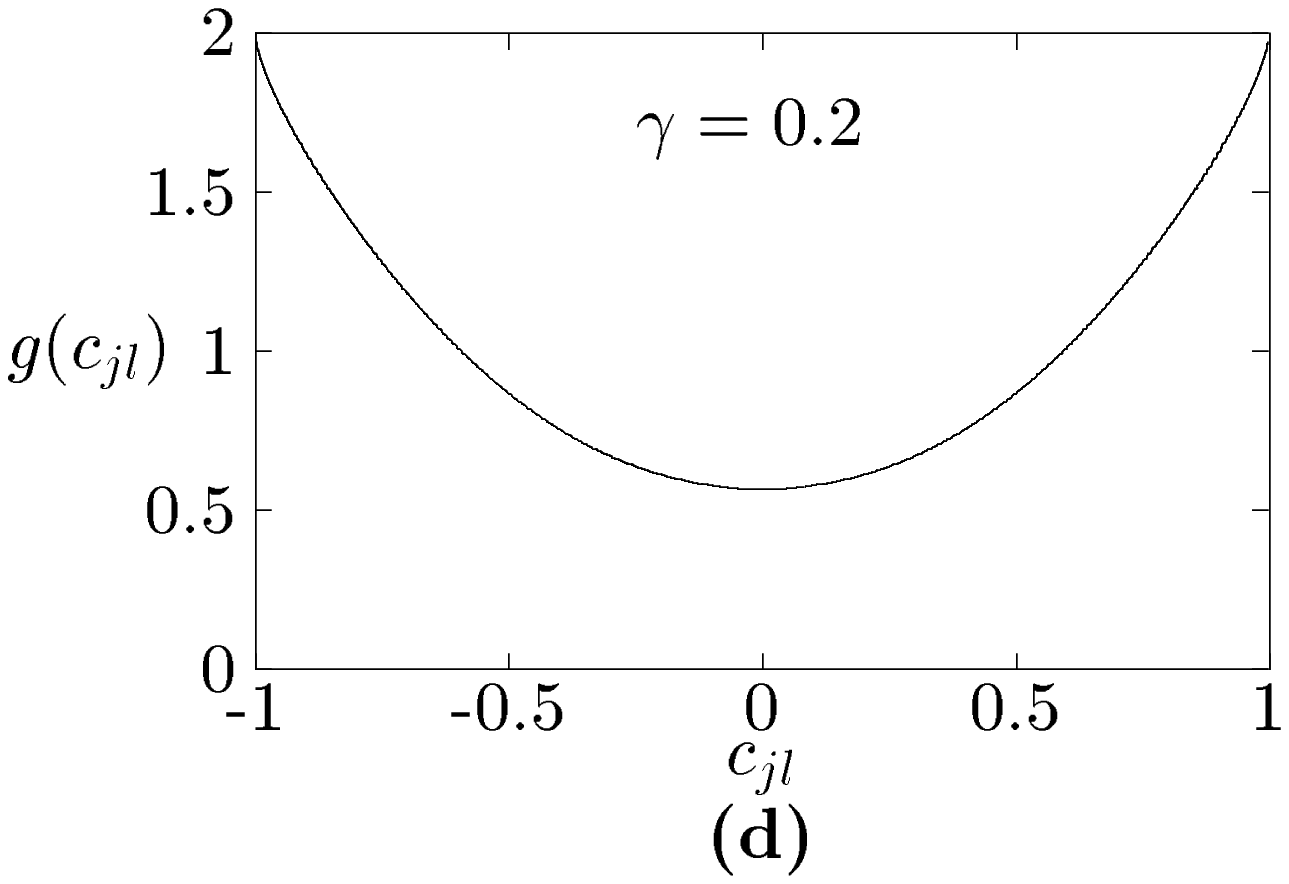}
\caption{The nonlinear coefficient $g(c_{jl})$ of the standing wave amplitude 
equation as a function of $c_{jl}$ for gravity-capillary waves of 
$\Gamma_0=1/3$ 
with the linear damping coefficient $\gamma=0.02$ in (a), $\gamma=0.05$ in 
(b), 
$\gamma=0.1$ in (c), and $\gamma=0.2$ in (d).}
\label{fig-gTheta13}
\end{figure}

\begin{figure}
\vspace{5in}
\includegraphics{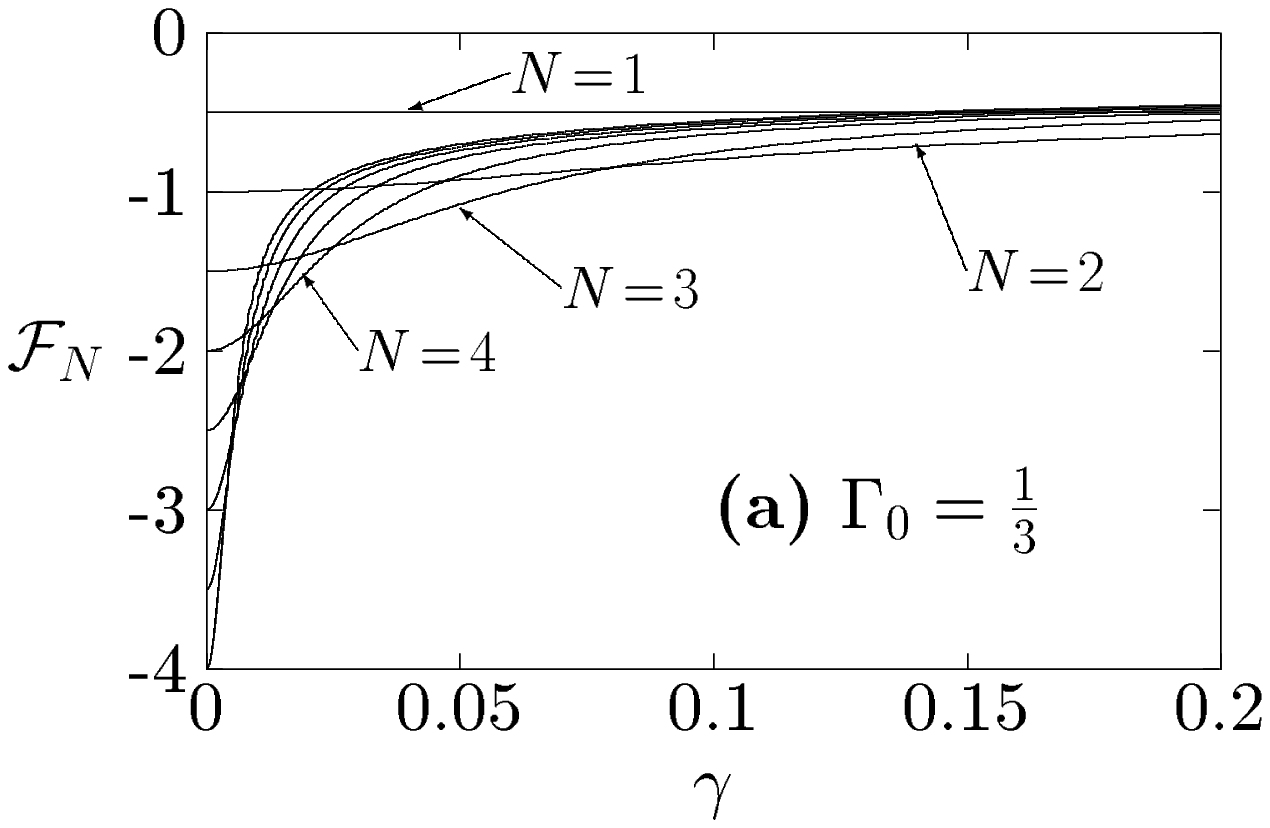}
\includegraphics{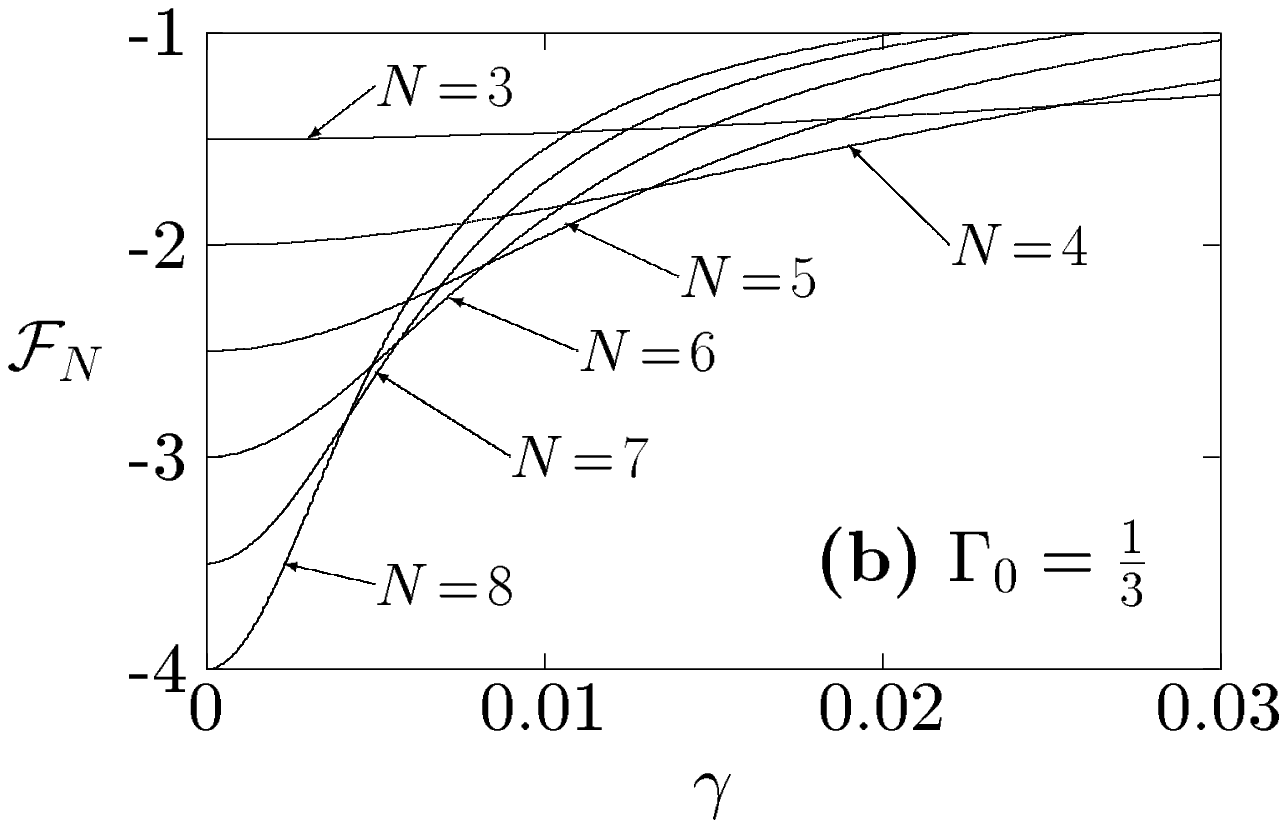}
\includegraphics{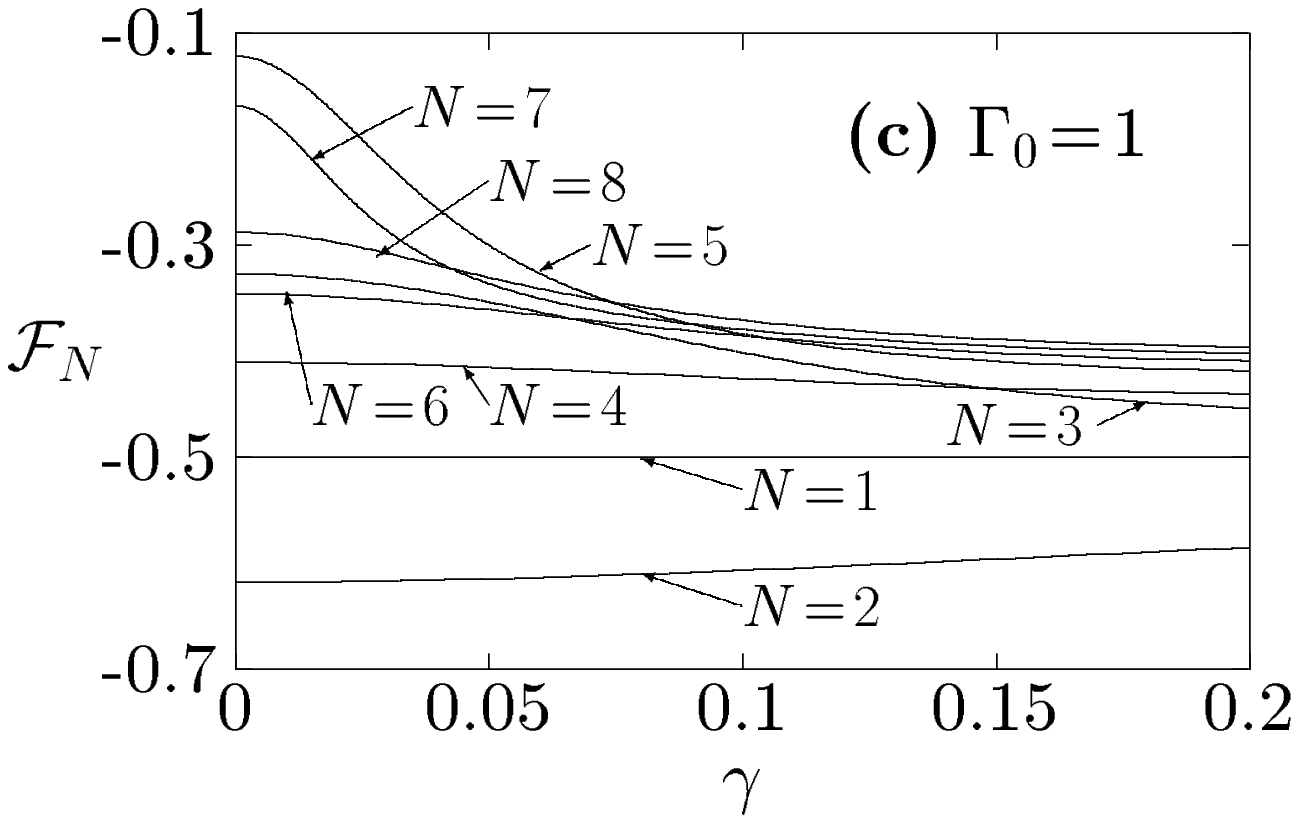}
\includegraphics{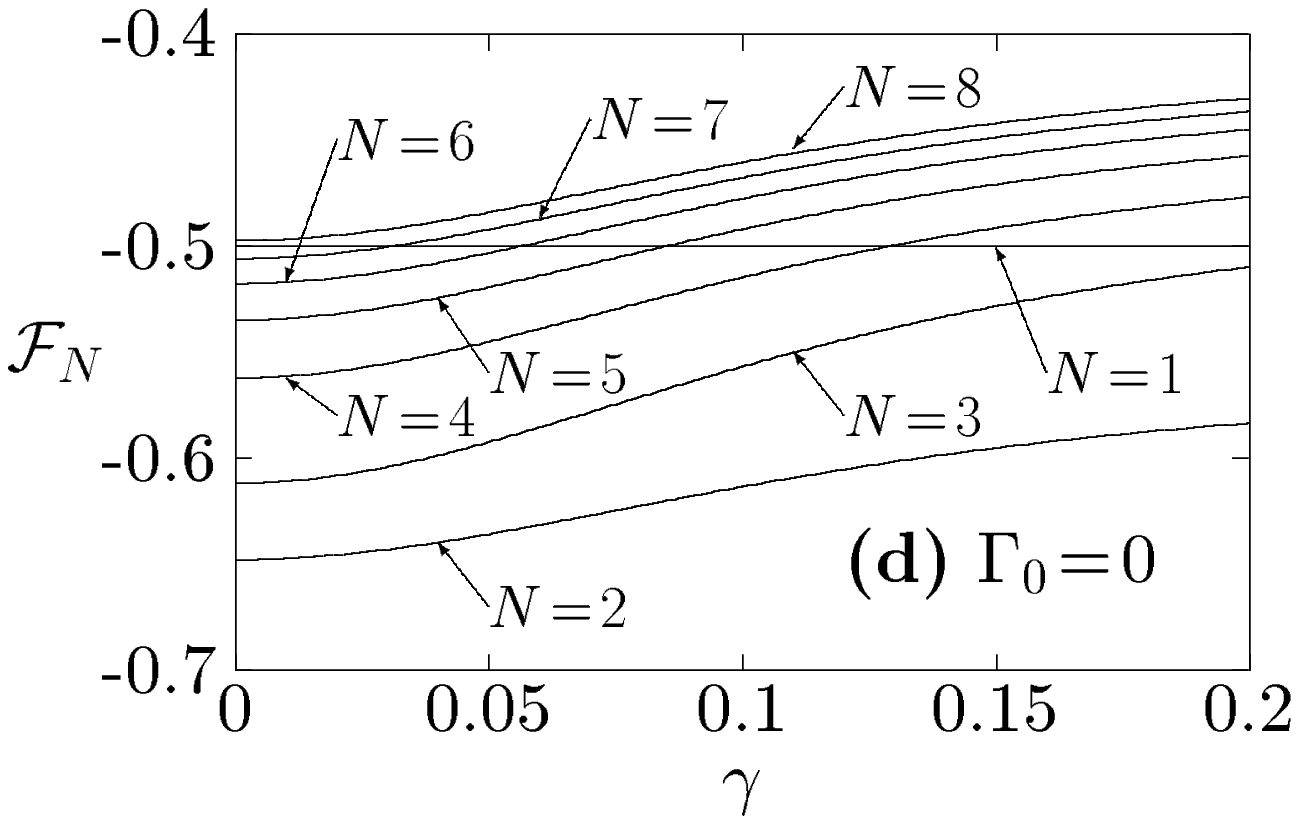}
\caption{The values of the Lyapunov function $\sF_N$ ($N=1,2,3,4,5,6,7,8$) 
as a function of $\gamma$ for $\Gamma_0 = 1/3$ in (a), $\Gamma_0 =1$ (capillary
waves) in (c), and $\Gamma_0 = 0$ (gravity waves) in (d).  The portion of (a)
with small values of $\gamma$ is shown in (b).}
\label{fig-Lyapunov13}
\end{figure}

We summarize our results concerning regular patterns 
in Fig.~\ref{fig-Lyapunov13}.
We present the values of the Lyapunov function $\sF_N$ as a function of 
$\gamma$
for $N=1, 2, 3, 4, 5, 6, 7, 8$, and $\Gamma_0 = 1/3$, 1, and 0.  
For $\Gamma_0 = 1$ (Fig.~\ref{fig-Lyapunov13}(c)) and 0
((Fig.~\ref{fig-Lyapunov13}(d)), patterns of square symmetry ($N=2$) have the
lowest values of $\sF_N$ for all values of $\gamma < 0.2$, whereas the system 
favors patterns of different symmetries in different ranges of the value for
$\gamma$ for $\Gamma_0 = 1/3$.  The second-harmonic resonance for $\Gamma_0 = 
1/3$, 
becomes less damped as $\gamma$ decreases, and thus the value of the 
self-interaction nonlinear coefficient $g(0)$ becomes larger (see
Fig.~\ref{fig-gTheta13}).  In other words, the curve $g(c_{jl})$ has 
a wider flat center region and a sharper increase near $c_{jl} = \pm 1$, and 
therefore pattern structures with larger $N$ are favored.
Table \ref{tab-Lyaponov_favored}
shows the favored structures, and the corresponding ranges of $\gamma$.  

A couple of comments on the various patterns discussed above are in order.  
(i) A two-dimensional regular pattern structure with the spatial form 
$\sum_{j=1}^N\left(A_j\exp(i\hkj\cdot\vx) + c.c.\right) $ 
has $N$ degrees of freedom.  These $N$ degrees of freedom appear as the phase 
of
the complex amplitude $A_j = A_0\exp(\theta_j)$ for $j=1,\cdots, N$.
Among them, two correspond to spatial translations, whereas the
other $N-2$ degrees of freedom represents the phason modes 
(Golubitsky, Swift \& Knobloch 1984; 
Malomed, Nepomnyashchi\u{i} \& Tribelski\u{i} (1989)).  
For $N=3$, the phase degeneracy for the 
single phason mode, which corresponds to hexagonal or triangular states, can 
be lifted by higher order nonlinear terms (Golubitsky, Swift \& Knobloch 1984; 
Muller 1993).  Although 
it is beyond the scope of this article, it will be interesting to see how 
the phase relations of a spatial pattern is determined by  higher order 
nonlinear
terms.  (ii) We predict that quasipatterns with large values of $N$ occur at 
very small values of the linear damping coefficient $\gamma$.  As discussed 
in the introduction, for very small values of $\gamma$ the mode 
quantization effect can be quite severe for finite-size systems.  This implies 
that experimental verification of the quasipatterns with large values of $N$ 
can be difficult \footnote{In laboratory experiments, mode quantization also
depends on the nature of boundary conditions.  Certain ``soft" boundary 
conditions may relax the strict quantization requirements and allows access 
to the larger system regime than the actual size of the system
(Douady 1990; Bechhoefer {\em et al.} 1995).}.  
Obviously, a similar problem appears in numerical solutions of Faraday waves.

The linear stability of solutions (Eq.~\ref{eq:SWA_solution}) of the standing
wave amplitude equations (Eq.~(\ref{eq:SWA2})) can be determined by the 
spectrum 
of growth rates $\sigma$ of amplitude perturbations $A_j \propto  
e^{\sigma t}$ since all phase perturbations are neutrally stable as is 
obvious from Eq.~(\ref{eq:SWA2}).  Equivalently, the linear stability of
solutions can also be obtained by the eigenvalue spectrum of the matrix 
$\partial^2\sF/(\partial |A_j|\partial |A_l|)$, linearized around the 
stationary 
solutions.  In the context of the standing wave amplitude equations
(Eq.~\ref{eq:SWA_solution}), we can only consider very limited set of
perturbations.  The usefulness of the stability analysis is to identify 
unstable
solutions.  A stable solution in the context of the standing wave amplitude 
equations can, however, be unstable to other perturbations, such as a 
traveling 
wave mode or a perturbation with spatial variations.  In the rest of this
subsection, stable or unstable solutions are with respect to the perturbations
that are allowed by the SWAE's. 

Let us consider a stationary solution with $|A_j| = a_0$ given by
Eq.~(\ref{eq:SWA_solution}) for $j=1, \cdots, N$ and $|A_j| = 0$ for
$j=N+1,\cdots,M$ ($M \ge N$), and small real perturbations $b_j$ to $|A_j|$ 
for 
$j=1, \cdots, M$.  On substituting the stationary solution with the 
perturbations
into Eq.~(\ref{eq:SWA2}), we obtain the following linearized equations for 
$b_j$:
$$ 
\frac{1}{\gamma}\partial_T b_j = 
\left[1-a_0^2\left(3+\sum_{l=1(l\ne j)}^{N} g(c_{jl})\right)\right] b_j
- 2a_0^2 \sum_{l=1(l\ne j)}^{N} g(c_{jl})b_l, 
$$ 
for $j=1, \cdots, N$, and
$$
\frac{1}{\gamma}\partial_T b_j = \left(1 
               - a_0^2\sum_{l=1}^{N}g(c_{jl})\right)b_j, 
$$ 
for $j=N+1, \cdots, M$.  

\begin{figure}
\vspace{5.0in}
\includegraphics{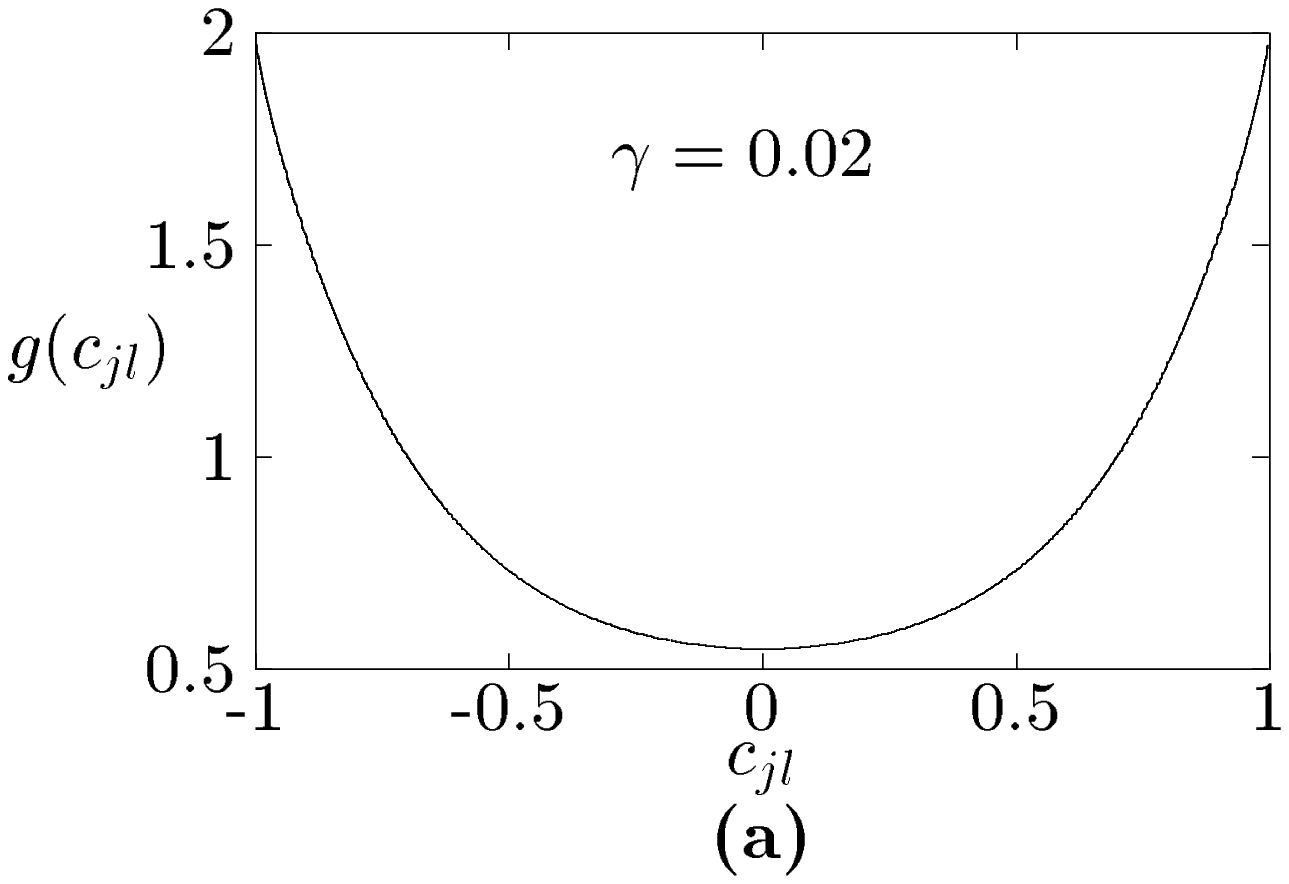}
\includegraphics{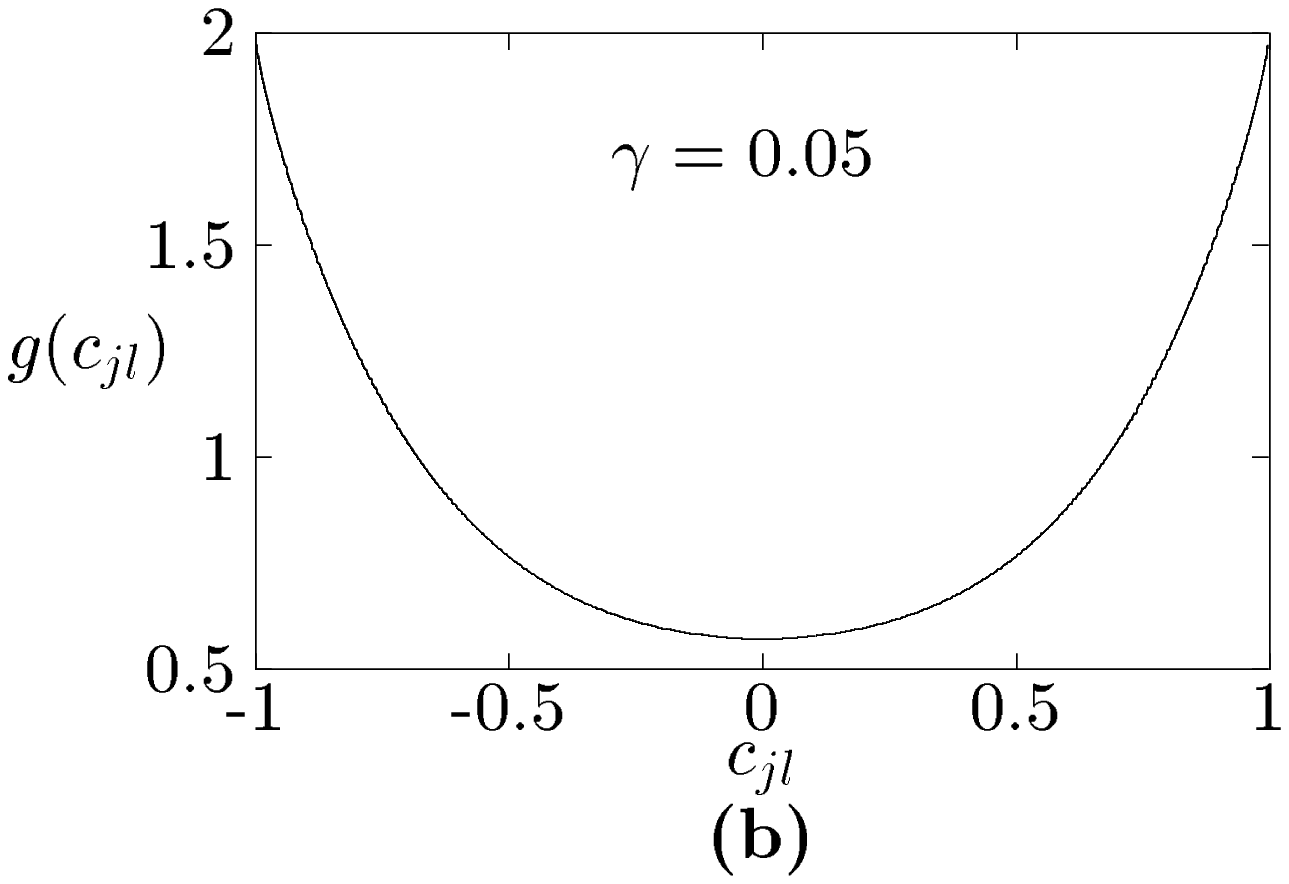}
\includegraphics{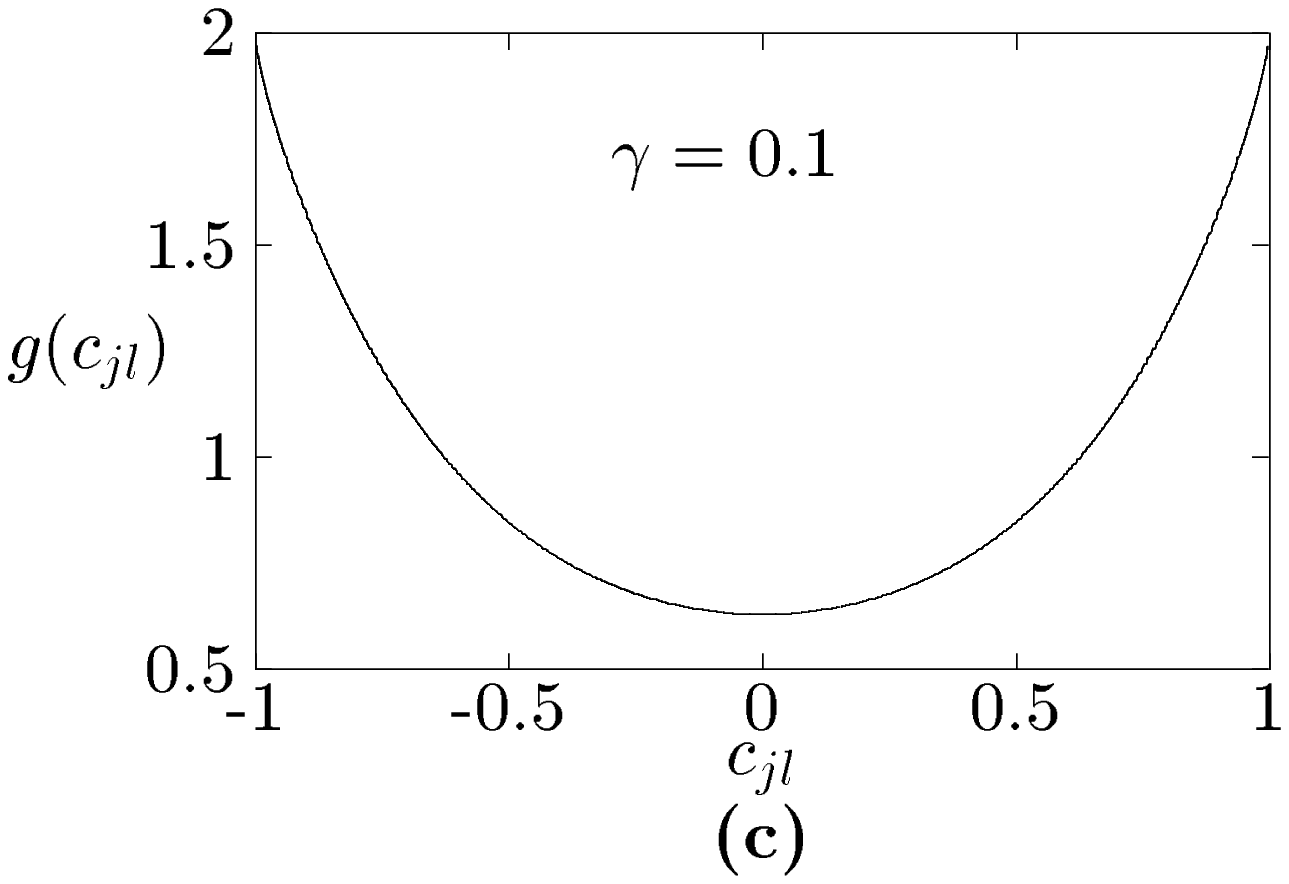}
\includegraphics{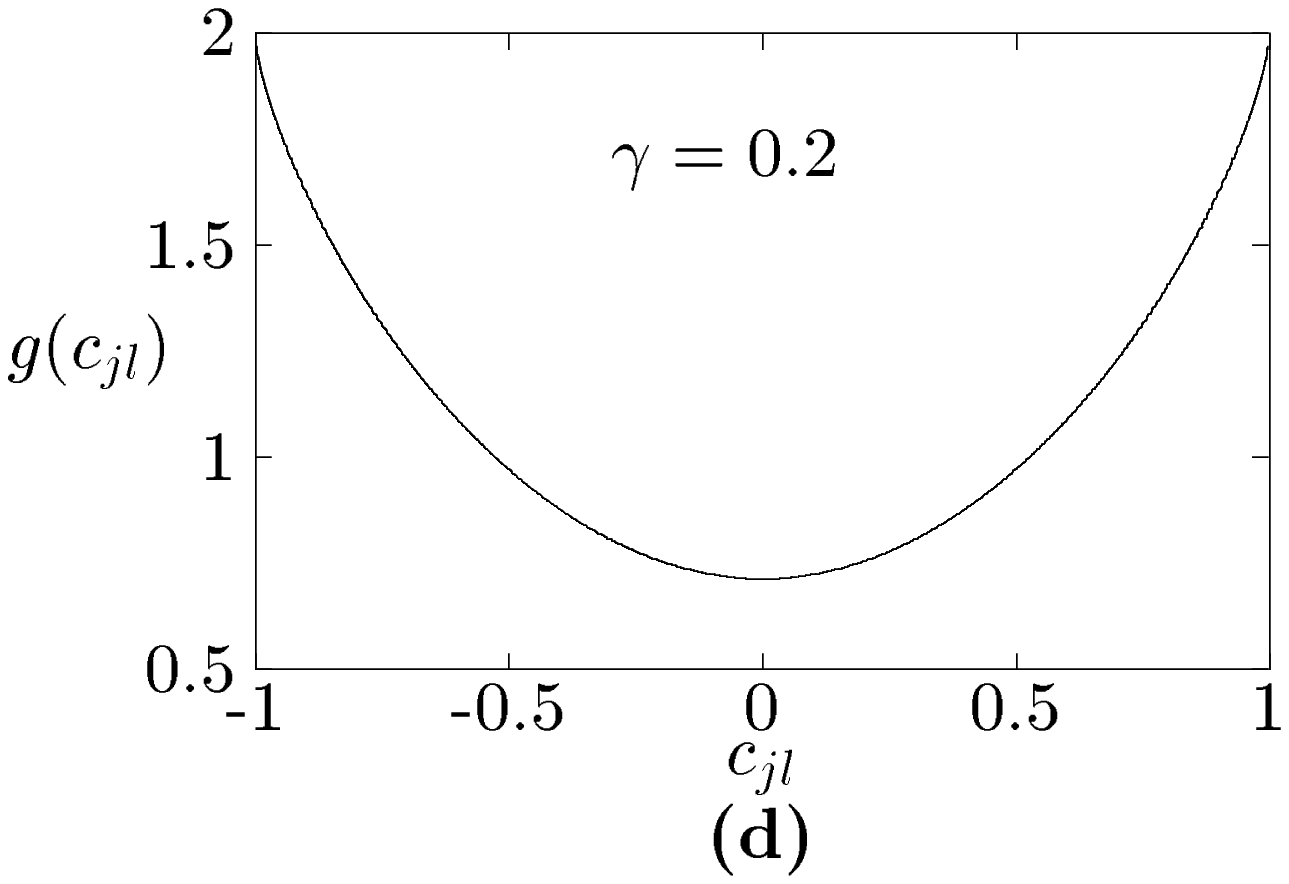}
\caption{The nonlinear coefficient $g(c_{jl})$ of the standing wave amplitude
equation as a function of $c_{jl}$ for purely gravity waves ($\Gamma_0=0$) 
with the linear damping coefficient $\gamma=0.02$ in (a), $\gamma=0.05$ in 
(b), $\gamma=0.1$ in (c), and $\gamma=0.2$ in (d).}
\label{fig-gTheta0.0}
\end{figure}

For parallel roll stationary solutions ($N=1$), we have $a_0 = 1$ and,
\bea
\frac{1}{\gamma}\partial_T b_1 = -2b_1,
\nonumber \\
\frac{1}{\gamma}\partial_T b_j = (1-g(c_{1j}))b_j,
\nonumber
\eea  
for $j=2, \cdots, M$.  Thus if $g(c_{1j}) < 1$ for some values of $c_{1j}$,
the parallel roll stationary solution is unstable. From the $g(c_{jl})$ 
curves 
in Figures \ref{fig-gTheta1.0}, \ref{fig-gTheta0.0}, and \ref{fig-gTheta13}, 
we conclude that parallel roll stationary solutions for Faraday waves in 
weakly 
dissipative fluids are not stable.

The linear stability of square patterns ($N=2$ and $a_0 = 1/\sqrt{1+g(0)}$) is 
determined from the following linear system,
\bea
\frac{1+g(0)}{2\gamma}\partial_T b_1 = -b_1 - g(0)b_2,
\nonumber \\
\frac{1+g(0)}{2\gamma}\partial_T b_2 = -b_2 - g(0)b_1,
\nonumber \\
\frac{1}{\gamma}\partial_T b_j = \left(1-\frac{g(c_{1j})+g(c_{2j})}{1+g(0)}
\right) b_j,
\nonumber
\eea 
for $j=3, \cdots, M$.  Thus if $-1 > g(0) > 1$ and
$$ 
\lambda_{sq} = 1-\frac{g(c_{1j}) + g(c_{2j})}{1+g(0)} < 0,
$$ 
square patterns would be stable.  The first condition is satisfied as 
shown in
Figures \ref{fig-gTheta1.0}, \ref{fig-gTheta0.0}, and \ref{fig-gTheta13}.
Since $|c_{2j}| = \sqrt{1-c_{1j}^2}$, we have
$$ 
\lambda_{sq}(c_{1j}) = 1-\frac{g(c_{1j}) + g(\sqrt{1-c_{1j}^2})}{1+g(0)},
$$ 
where $-1 \le c_{1j} \le 1$. The growth rate $\lambda_{sq}$ 
of a standing wave perturbation in an arbitrary direction to the 
stationary square solution is always negative except for $\Gamma_0=1/3$
and small values of damping coefficient $\gamma = 0.02$.  Interestingly,
this is the parameter regime for $\Gamma_0$ and $\gamma$ where 
hexagonal/triangular patterns are found to have lower values of the Lyapunov
function than square patterns.  The results of the stability analysis tell
us that in this parameter range square patterns are in fact unstable, 
and thus correspond to a local maximum or a saddle point of the 
Lyapunov function.  This result in turn guarantees that square patterns
will not be seen in Faraday waves for this parameter range.  The
exact parameter range, however, cannot be determined from the stability
analysis since we are only considering very restricted form of perturbations
in the context of the standing wave amplitude equations.  

The linear stability of hexagonal or triangular standing wave patterns ($N=3$ 
and $a_0 = 1/\sqrt{1+2g(1/2)}$) can be determined from the following linear 
system,
\bea
\frac{1+2g(1/2)}{2\gamma}\partial_T b_1 = -b_1 - g(1/2) (b_2 + b_3),
\nonumber \\
\frac{1+2g(1/2)}{2\gamma}\partial_T b_2 = -b_2 - g(1/2) (b_1 + b_3),
\nonumber \\
\frac{1+2g(1/2)}{2\gamma}\partial_T b_3 = -b_3 - g(1/2) (b_1 + b_2), 
\nonumber \\
\frac{1}{\gamma}\partial_T b_j = \left(1-\frac{g(c_{1j}) + g(c_{2j}) + 
g(c_{3j})}
                                              {1+2g(1/2)}\right)b_j,
\nonumber 
\eea
for $j=4, \cdots, M$.  Therefore, hexagonal or triangular
patterns are unstable if $g(1/2) < -1/2$, or $g(1/2) > 1$ or,
$$ 
\lambda_{ht} \equiv 1-\frac{g(c_{1j}) + g(c_{2j}) + g(c_{3j})}{1+2g(1/2)} > 0,
$$ 
for $j=4, \cdots, M$.  Similar stability analyses can also be performed for 
quasipatterns ($N \ge 4$).

\subsection{Envelope Equations}
\label{sec:envelope-sinusoidal}

We have assumed so far that Faraday wave patterns consist of a
set of spatially uniform (i.e., no modulations) standing waves although we
have considered the amplitude of the each standing wave to be slowly varying 
in time.  It is conceivable, however, that the standing wave amplitudes
may also have slow spatial variations, i.e, standing waves with spatially
modulated amplitudes.  In fact, the slow spatial variations of the amplitudes
can be nicely incorporated into the amplitude equations within the framework 
first discussed by \cite{re:newell69} for pattern formation
in Rayleigh-B\'enard convection.

Amplitude equations including slow spatial variations of the amplitudes 
are often called envelope equations.  In principle one can also derive
a set of coupled standing wave envelope equations for parametric surface 
waves from the hydrodynamic equations with the assumption that spatial
variations of the amplitudes are of a slow scale of the order of
$\varepsilon^{1/2}$, i.e., $\vX = \varepsilon^{1/2}\vx$ and $A_j = A_j(\vX, 
T)$.
With these assumptions, there will be no nonlinear terms involving spatial
derivatives in the envelope equations up to ${\cal O}(\varepsilon^{3/2})$ 
\footnote{Nonlinear terms involving spatial derivatives would appear in the
envelope equations if the standing wave amplitude equations had quadratic
nonlinear terms (Brandt 1989).}  As a result, only linear terms involving
spatial derivatives will appear in standing wave envelope equations (SWAE's), 
and the cubic nonlinear terms will be exactly the same as the standing wave 
amplitude equations (Eq.~(\ref{eq:SWA1})) (SWAE's).  The linear terms 
involving 
spatial derivatives can be obtained by a similar perturbative expansion as 
for the SWAE's as well as from symmetry and invariance arguments, and the 
growth
rate of the linearly unstable modes (Newell 1974; Cross \& Hohenberg 1993).
We have chosen the latter way, which is much simpler.

The possible scalar terms up to ${\cal O}(\varepsilon^{3/2})$ are 
$\left(\hkj\cdot\nabla_{X}\right)A_j$, $\nabla_{X}^2A_j$, and 
$\left(\hkj\cdot\nabla_{X}\right)^2A_j$.  Since the envelope equation for the 
standing wave amplitude $A_j$ is expected to be invariant under the
transformation $\hkj \rightarrow -\hkj$, the first term will not appear in 
the equation.  The second term implies the same form of the transverse and
longitudinal variations for a standing wave amplitude $A_j$.  Since a standing 
wave pattern breaks the rotational symmetry of an isotropic two-dimensional 
system, the transverse and longitudinal spatial variations are qualitatively 
different, and thus the second term also cannot appear in the envelope 
equations
\footnote{With the assumption of different slow spatial scales for the
transverse and longitudinal modulations in analogy to Rayleigh-B\'enard 
convection, scalar terms 
of the following form are permissible in standing wave amplitude equations:
$\left[\hkj\cdot\nabla_{X} 
       -(i/2)\left(\hkj^{\perp}\cdot\nabla_{X}\right)^2\right]^2A_j$ with
$\hkj^{\perp}$ a unit vector perpendicular to $\hkj$. We do not consider 
such terms here
since we are only computing the coherence length $\xi_{0}$ in the longitudinal
direction. The more general case of transverse modulations has been 
explicitly addressed by \cite{re:zhang94} .}.  
Therefore, we only need to consider the last term 
$\left(\hkj\cdot\nabla_{X}\right)^2A_j$ in the envelope equations,
\be
\frac{1}{\gamma}\frac{\partial A_j}{\partial T} =  A_j
+ \xi_0^2 \left(\hkj\cdot\nabla_{X}\right)^2A_j(\vX, T)
- \left[|A_j|^2 +\sum_{l=1(l\ne j)}^{N} g(c_{jl})
|A_l|^2 \right] A_j,
\ee
where $\xi_0$ is determined from the growth rate of the linearly unstable 
eigenmode as given by Eq.~(\ref{eq:growth_rate}),
$$ 
\sigma_{+}(k)=-\gamma+\sqrt{f^2-\left[(k-1)(G_0+3\Gamma_0)/2\right]^2}.
$$ 
From the general arguments of \cite{re:newell74} and 
\cite{re:cross93}, we expand the growth rate $\sigma_{+}(k)$ around the 
critical
wavenumber $k=1$ as
$$ 
\sigma_{+}(k) = \frac{1}{\tau_0} \left[\varepsilon - \xi_0^2(k-1)^2\right] + 
\cdots.
$$ 
Hence we find that $\tau_0 = 1/\gamma$ and
\be
\xi_0^2 = -\frac{\tau_0}{2}
          \left(\frac{\partial^2\sigma_{+}(k)}{\partial k^2}\right)_{k=1}
        = \frac{\tau_0}{2f}\left(\frac{G_0+3\Gamma_0}{2}\right)^2,
\ee
The envelope equations now can be written as, 
\be
\label{eq:SWA_envelope}
\frac{1}{\gamma}\frac{\partial A_j}{\partial T} =  A_j
+ \left(\frac{G_0+3\Gamma_0}{2\sqrt{2}\gamma}\right)^2
        \left(\hkj\cdot\nabla_{X}\right)^2 A_j
- \left[|A_j|^2 +\sum_{l=1(l\ne j)}^{N} g(c_{jl})
|A_l|^2 \right] A_j,
\ee
for $j = 1, \cdots, N$.  We have set $f=\gamma$ in the spatial derivative term
since the difference is of higher order.  

The envelope equations (Eq.~(\ref{eq:SWA_envelope})) are also of gradient form
$\frac{1}{\gamma}\frac{\partial A_j}{\partial T} = -\delta \sL/\delta A_j^*$, 
with a Lyapunov functional
\bea
\sL = \int d\vx 
\Biggl[-\sum_{j=1}^{N} |A_j|^2 + 
      \left|\frac{G_0+3\Gamma_0}{2\sqrt{2}\gamma}
      \left(\hkj\cdot\nabla_{X}\right)A_j\right|^2
\nonumber \\
\label{eq:Lyapunov_functional}
      + \frac{1}{2}\sum_{j=1}^{N} |A_j|^2
        \left(|A_j|^2 + \sum_{l=1(l\ne j)}^N g(c_{jl}) |A_l|^2\right)\Biggr].
\eea

An important point about the standing wave envelope equations 
(Eq.~\ref{eq:SWA_envelope}) and the Lyapunov functional 
(Eq.~\ref{eq:Lyapunov_functional}) is that the coefficient
of the spatial derivative term is much greater than one, i.e.,
\be
\label{eq:coherence_length}
\xi_0 = \frac{G_0+3\Gamma_0}{2\sqrt{2}\gamma} \gg 1,
\ee
since $\gamma \ll 1$ for weakly dissipative surface waves.  Physically $\xi_0$ 
is 
a measure of the coherence length of the wave pattern.  Thus
Eq.~(\ref{eq:coherence_length}) implies standing wave patterns of Faraday 
waves in weakly dissipative fluids have a coherence length much longer than 
their wavelength.  This should be compared to Rayleigh-B\'enard convection, 
where we
have $\xi_0 \sim 1$ (Newell \& Whitehead 1969; Cross \& Hohenberg 1993), 
and to directional solidification of a binary alloy, 
where $\xi_0 \ll 1$ due to the extremely flat neutral stability curve 
(Mullins \& Sekerka 1964).  Because $\xi_0 \gg 1$, highly ordered square
patterns as large as $30-40$ wavelengths are often observed in Faraday waves 
in weakly viscous fluids (Gaponov-Gerkhov \& Rabinovich 1990).  

\section{Summary and Discussion}
\label{sec:conclusions}

We have studied in this chapter pattern formation in weakly damped Faraday 
waves
by deriving a set of linear damping quasi-potential equations and by performing
a multiscale asymptotic expansion close to onset. Standing wave equations have 
been derived that explicitly incorporate higher harmonic terms in the linear 
neutral solutions. These terms are seen to be important for the saturation 
of the wave amplitude (amplitude-limiting effect by the driving force).  
We have also studied in detail the effect of triad resonant interactions among 
capillary-gravity waves on nonlinear pattern selection.  Triad resonant
interactions are found to be the main reason for the appearance of square
patterns in capillarity dominated  Faraday waves.  By increasing the gravity 
wave component,
the triad resonant condition is altered.  As a result, square patterns can
become unstable, and hexagonal or quasicrystalline patterns can be stabilized.

The importance of higher harmonic terms in the linear neutral solutions
also has implications for parametric surface waves in highly viscous fluids.
Because of the larger threshold values of the driving force for highly
viscous fluids, higher harmonic terms in the linear neutral solutions can 
be even more important than in the weakly damped case considered here.  More 
than one higher harmonic term can be important in that case.

Our results provide justification for the observed selected patterns of
square symmetry near onset in fluids of low viscosity
(Lang 1962; Ezerskii, Korotin \& Rabinovich
1985; Tufillaro, Ramshankar \& Gollub 1989; Ciliberto, Douady \& Fauve
1991; Bosch \& van de Water 1993; Edwards \& Fauve 1993; M\"uller 1993). 
Without requiring any additional assumptions, Faraday waves close to onset
are potential, and minimization of the associated Lyapunov functional leads to
square patterns in the capillarity dominated regime. This is in agreement with
most experimental studies, with the exception of the work by
\cite{re:christiansen95} who observed quasi-patterns where we predict squares.
More recently, a systematic survey of pattern
selection in a system of large aspect ratio by Kudrolli \& Gollub (1996)
has again documented the transition to squares for capillary dominated
waves, but also shown the transition to hexagons in the vicinity of 
$\Gamma = 1/3$, in agreement with our predictions. In the range in which
we predict stable quasi-patterns, Kudrolli \& Gollub (1996) observe hexagons
instead. The origin of this discrepancy remains to be investigated.

\begin{acknowledgments}
We are indebted to Maxi San Miguel for many useful discussions.
This work was supported by the Microgravity Science and Application
Division of the NASA under Contract No. NAG3-1284.
This work was also supported in part by the Supercomputer
Computations Research Institute, which is partially funded by the U.S.
Department of Energy, contract No. DE-FC05-85ER25000.
\end{acknowledgments}

\newpage

\begin{table}
\centering
\caption{Patterns with the lowest values of the Lyapunov function
and their corresponding ranges of $\gamma$.  These values correspond to mixed
capillary-gravity waves with $\Gamma_0 = 1/3$.}
  \begin{tabular}{cl}\hline
   range of $\gamma$ & favored pattern \\
  \hline
   $\gamma > 0.0809          $   & square patterns \\
   $0.0253 < \gamma < 0.0809 $   & hexagonal/triangular patterns \\
   $0.0132 < \gamma < 0.0253 $   & eightfold quasipatterns \\
   $0.0082 < \gamma < 0.0132 $   & tenfold quasipatterns \\
   $0.0057 < \gamma < 0.0082 $   & twelvefold quasipatterns \\
   $0.0042 < \gamma < 0.0057 $   & fourteenfold quasipatterns \\
  \hline 
  \end{tabular}
\label{tab-Lyaponov_favored}
\end{table}

\newpage

\newpage

\newpage

\newpage

\newpage

\newpage

\newpage

\newpage

\newpage

\end{document}